\documentclass[10pt,final,journal,a4paper,oneside,twocolumn]{IEEEtran}
\usepackage[pdftex]{graphicx}
\usepackage{amsmath,amsfonts,amssymb,amsthm}
\usepackage[makeroom]{cancel}
\usepackage{float}
\usepackage{cite}
\usepackage{booktabs}
\usepackage{multirow}

\newtheorem*{remark}{Remark}
\def\authorname{}
\def\ttitle{}
\newcommand*{\supervisor}[1]{\def\supname{#1}}
\newcommand*{\thesistitle}[1]{\def\@title{#1}\def\ttitle{#1}}

\newcommand*{\authortitlepage}[1]{\def\authorname{#1}}
\newcommand*{\university}[1]{\def\univname{#1}}
\newcommand*{\department}[1]{\def\deptname{#1}}
\newcommand*{\group}[1]{\def\groupname{#1}}
\newcommand*{\faculty}[1]{\def\facname{#1}}

\renewcommand{\IEEEQED}{\IEEEQEDopen}  % Change QED symbol to an open symbol

\thesistitle{{Feedforward Control of Magnetically Levitated Planar Actuators}} % Your thesis title, this is used in the title and abstract, print it elsewhere with \ttitle
\supervisor{{Ioannis~Proimadis MSc\\Yanin~Kasemsinsup MSc\\dr.ir. Roland~T\'oth}} % {Dr. James \textsc{Smith}} % Your supervisor's name, this is used in the title page, print it elsewhere with \supname
\authortitlepage{Tom Bloemers\\ ID: 0829970} % Your name, this is used in the title page and abstract, print it elsewhere with \authorname
\university{Eindhoven University of Technology} % Your university's name and URL, this is used in the title page and abstract, print it elsewhere with \univname
\department{{Electrical Engineering}} % Your department's name and URL, this is used in the title page and abstract, print it elsewhere with \deptname
\group{{Control Systems}} % Your research group's name and URL, this is used in the title page, print it elsewhere with \groupname
\faculty{{Electrical Engineering}} % Your faculty's name and URL, this is used in the title page and abstract, print it elsewhere with \facname

\begin{document}
% Titlepage
%\begin{titlepage}
%\begin{center}
%
%\textsc{\LARGE \univname}\\[1.5cm] % University name
%% \textsc{\Large Doctoral Thesis}\\[0.5cm] % Thesis type
%%
%\HRule \\[0.4cm] % Horizontal line
%{\huge \bfseries \ttitle}\\[0.4cm] % Thesis title
%\HRule \\[0.4cm] % Horizontal line
%\large Systems and Control \\[1cm]
%
%
%\begin{minipage}[t]{0.4\textwidth}
%\begin{flushleft} \large
%\emph{Author:}\\
%{\authorname} % Author name - remove the \href bracket to remove the link
%\end{flushleft}
%\end{minipage}
%\begin{minipage}[t]{0.4\textwidth}
%\begin{flushright} \large
%\emph{Supervisor:} \\
%{\supname} % Supervisor name - remove the \href bracket to remove the link
%\end{flushright}
%\end{minipage}\\[3cm]
%
%\vfill
%
%\large \groupname\\\deptname\\[2cm] % Research group name and department name
%
%{\large December 22, 2016}\\[3cm] % Date
%\includegraphics[totalheight=0.08\textheight]{Figures/TUE.eps} % University/department logo - uncomment to place it
%%
%\vfill
%\end{center}
%\end{titlepage}

% paper title
\title{\ttitle}
%
%
%
% author names and IEEE memberships
% note positions of commas and nonbreaking spaces ( ~ ) LaTeX will not break
% a structure at a ~ so this keeps an author's name from being broken across
% two lines.
% use \thanks{} to gain access to the first footnote area
% a separate \thanks must be used for each paragraph as LaTeX2e's \thanks
% was not built to handle multiple paragraphs
%
\author{Tom~Bloemers~ID:0829970 \\
{Supervisors}:~Ioannis~Proimadis,~Yanin~Kasemsinsup~and~Roland~T\'oth\\
Department~of~Electrical~Engineering, Control~Systems, Eindhoven~University~of~Technology}

% The paper headers
\markboth{Systems~and~Control:~Internship 22-12-2016}%
{Systems~and~Control:~Internship}
% The only time the second header will appear is for the odd numbered pages
% after the title page when using the twoside option.

% use for special paper notices
%\IEEEspecialpapernotice{(Invited Paper)}

% make the title area
\maketitle

% -------------------------------------------------------------------------------- Abstract --------------------------------------------------------------------------------
% As a general rule, do not put math, special symbols or citations
% in the abstract or keywords.
\begin{abstract}
The present report summarizes the work conducted during the internship on Feedforward Control of the Magnetic Levitation Setup. Different feedforward strategies, specifically tailored for this setup, are developed and reviewed. These feedforward methods explicitly take the intrinsic position-dependent behavior of the magnetic levitation setup into account. Additionally, closed-loop stability of the given setup is assessed. All investigations are carried out under the rigid-body assumption of the structure. Analysis and simulation show the potential performance improvement obtained with such feedforward strategies.
\end{abstract}

% Note that keywords are not normally used for peerreview papers.
%\begin{IEEEkeywords}
%IEEE, IEEEtran, journal, \LaTeX, paper, template.
%\end{IEEEkeywords}

% For peer review papers, you can put extra information on the cover
% page as needed:
% \ifCLASSOPTIONpeerreview
% \begin{center} \bfseries EDICS Category: 3-BBND \end{center}
% \fi
%
% For peerreview papers, this IEEEtran command inserts a page break and
% creates the second title. It will be ignored for other modes.
\IEEEpeerreviewmaketitle

% -------------------------------------------------------------------------------- Introduction --------------------------------------------------------------------------------
\section{Introduction}
% The very first letter is a 2 line initial drop letter followed
% by the rest of the first word in caps.
%
% form to use if the first word consists of a single letter:
% \IEEEPARstart{A}{demo} file is ....
%
% form to use if you need the single drop letter followed by
% normal text (unknown if ever used by the IEEE):
% \IEEEPARstart{A}{}demo file is ....
%
% Some journals put the first two words in caps:
% \IEEEPARstart{T}{his demo} file is ....
%
% Here we have the typical use of a "T" for an initial drop letter
% and "HIS" in caps to complete the first word.
%
% You must have at least 2 lines in the paragraph with the drop letter
% (should never be an issue)

% OLD INTRO
%\IEEEPARstart{I}{n} lithographic processes, a wafer stage carrying a silicon wafer moves in a scanning motion with high speed and acceleration to accurately position the wafer with respect to the optical components. These optical components expose a pattern on the wafer that makes up the integrated circuit design. The movement of the wafer stage is required to be in the nanometer accuracy range to support the small feature sizes of the integrated circuit patterns. Small feature sizes enable more functionality e.g., the increased capacity of memory chips or faster microprocessors \cite{butler:pos_control_lithographic}.\\

\IEEEPARstart{I}{n} lithographic processes, silicon wafers are exposed by optical components to create a pattern of the integrated circuit design. The smaller these patterns can be made, the more functionality the chip can possess such as increased capacity of memory chips or faster microprocessors \cite{butler:pos_control_lithographic}. The silicon wafers are carried by a wafer stage, moving in a scanning motion with high speed and acceleration to accurately position the wafer with respect to the optical components. To enable the design of small patterns on the wafer, the movement of the wafer stage is required to be within the nanometer accuracy range.

A moving-magnet planar actuator was designed at Eindhoven University of Technology \cite{NAPAS:design}, \cite{Rovers:NAPAS_modeling}, under the acronym NAPAS (Nanometer-accurate planar actuation system). This system is a prototype for next generation wafer stages. To achieve high accelerations and accuracies, accurate modeling and control of the mechanical behavior of the translator is a requisite. To realize such a control solution Linear Time-Invariant (LTI) methods can possibly be restrictive in the light of the position dependent nature of the setup \cite{white_paper_ioannis}. The Linear Parameter-Varying (LPV) framework is an attractive alternative to incorporate the position-dependent behavior emerging in the input and output map of the translator. The LPV framework offers a linear input-to-output mapping, which depends on the so-called scheduling variables. These scheduling variables can be chosen such that they capture the position dependent behavior of the translator. LPV control methods are based on the extension of well-developed control tools for LTI systems. These control tools are extended such that they cope with the dependence on the scheduling variables while retaining the simplicity of LTI control tools.

To obtain nanometer-accurate positioning performance, feedforward and feedback-control play key roles. In general, feedforward is used to realize the required input 1) to follow a desired known reference signal in terms of output response, 2) to attenuate the effect of known disturbances, e.g., gravity. Feedback is used to stabilize the system, reject disturbances and to account for model uncertainties or unmodeled dynamics. During this internship, the focus lies on feedforward control of the NAPAS setup. In motion systems, it is typical to use model-based feedforward strategies such as \cite{yanin_exact_inverse}, \cite{trajectory_planning} and \cite{ff_motion_systems} or data-driven approaches such as \cite{data_based_feedforward}. An extensive review of feedforward strategies for LTI system representations can be found in \cite{FF_review}. In current wafer stages, strategies such as mass, jerk or snap feedforward \cite{jerk_ff}, \cite{MIMO_jerk_ff} are typically used to account for the rigid body dynamics. Due to the close resemblance to robotic systems, approaches based on the nonlinear dynamical equations such as in \cite{robotics_book} and \cite{robotics_book_2} can be also attractive. In this report, feedforward strategies specifically tailored for the NAPAS setup will be investigated. These methods are designed such that the position dependent behavior of the setup is taken into account. The advantages of these methods over traditional LTI feedforward strategies are shown through simulations.

This report begins with an overview of the rigid body dynamics of the NAPAS setup in Section \ref{section:system_dynamics}. Then, Section \ref{section:ff_methods} discusses the design aspects of the feedforward methods that are taken into consideration. Stability of the NAPAS setup is important, to prevent possible damage to the setup. Therefore, a closed-loop stability analysis is given in Section \ref{section:stability_analysis}. The feedforward controllers are assessed through simulations in Section \ref{section:simulations}. Finally, conclusions and recommendations are drawn in Section \ref{section:conclusions}.

% ------------------------------------------------------------------------ Rigid Body Dynamics ------------------------------------------------------------------------
\section{Rigid body dynamics}
\label{section:system_dynamics}
\begin{figure}[h]
\centering
\includegraphics[width=0.75\linewidth]{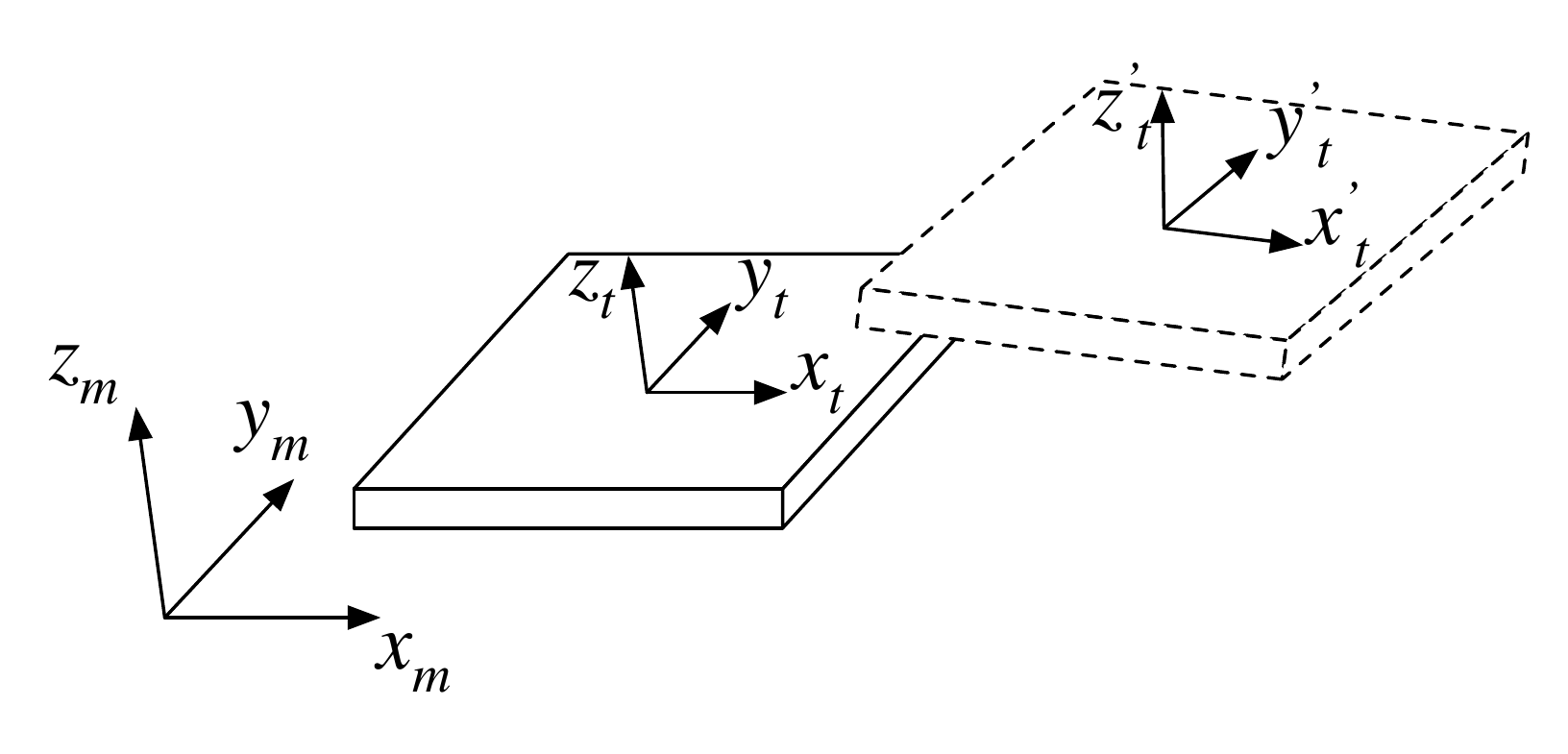}
\caption{Schematic of the NAPAS translator. The triplet $(x_m, y_m, z_m)$ denotes the global reference frame and $(x_t,y_t,z_t)$ denotes the translators reference frame. Let $(x^\prime_t, y^\prime_t, z^\prime_t)$ denote another position of the translator in the global reference frame.}
\label{fig:Translator}
\end{figure}

\begin{figure}[h]
\centering
\includegraphics[width=0.35\linewidth]{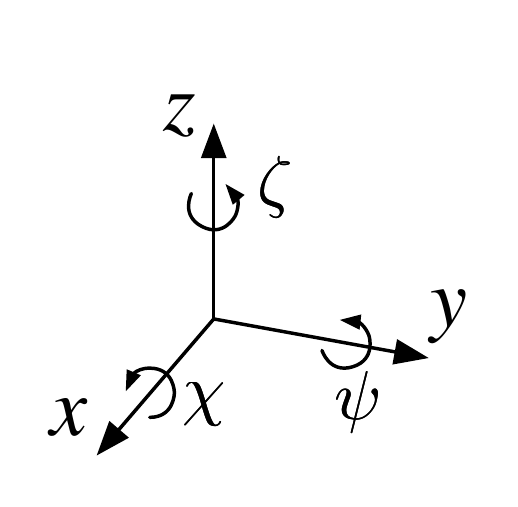}
\caption{The angles $(\chi,\psi,\zeta)$ used in the Pitch-Yaw-Roll representation.}
\label{fig:PYR}
\end{figure}

In this section the equations of motion for the NAPAS setup are discussed. The equations of motion are derived based on the Euler-Lagrange approach. Such an approach uses energy related considerations for the derivation of the equations of motion. In order to derive the rigid body dynamics in terms of a global reference frame, the kinetic and potential energy of a point on the translator have to be expressed with respect to this reference frame. A schematic diagram of the translator is shown in Figure \ref{fig:Translator}, where $(x_m,y_m,z_m)$ denotes the global reference frame and $(x_t,y_t,z_t)$ denotes the reference frame of the translator.
Let the position of the translator in the global reference frame be given by the generalized coordinates $q \in \mathbb{R}^{n_q}$ defined as
\begin{align}
q := \begin{bmatrix} x_s & y_s & z_s & \chi_s & \psi_s & \zeta_s \end{bmatrix}^\top,
\end{align}
where $(x_s,y_s,z_s)$ denote the translation of the Center of Mass (CoM) of the translator with respect to the global reference frame $(x_m,y_m,z_m)$ and the angles $(\chi_s,\psi_s,\zeta_s)$ denote the rotation of the center of mass around the axes of the global reference frame, as defined under the Pitch-Yaw-Roll representation \cite{robotics_book}. In this representation, the three angles, as in Figure \ref{fig:PYR}, denote the successive rotation around the three inertial frames: first a rotation $\chi$ around the $x$-axis (yaw), then a rotation $\psi$ around the y-axis (pitch) and finally a rotation $\zeta$ around the $z$-axis (roll).

The solution to the Euler-Lagrange equations lead to the equations of motion \cite{NAPAS_rigid_body}:
\begin{align}
\label{eqn:nlmodel}
M(q) \ddot{q} + C(q,\dot{q})\dot{q} &= W - D\dot{q},
\end{align}
where $M(q) \in \mathbb{R}^{n_q \times n_q}$ is the mass-matrix which is a positive definite matrix in the range of operation (proof in Appendix A), $C(q,\dot{q}) \in \mathbb{R}^{n_q \times n_q}$ is the Coriolis matrix and $D \in \mathbb{R}^{n_q \times n_q}$ contains the friction (dissipation) terms. $W \in \mathbb{R}^{n_w}$ represents the generalized forces and torques acting on the plate around the center of mass expressed in the global reference frame $(x_m,y_m,z_m)$:
\begin{align}
W &:= \begin{bmatrix} F_{x} & F_{y} & F_{z} & \tau_\chi & \tau_\psi & \tau_\zeta \end{bmatrix}^\top.
\end{align}
The matrices $M(q)$, $C(q,\dot{q})$ and $D$ are given in Appendix A. Important properties of the system description are:
\begin{itemize}
\item $M(q)$ is a symmetric and positive definite matrix in the range of operation (proof in Appendix \ref{appendix:system_matrices}).
\item $\dot{M}(q)-2C(q,\dot{q})$ is a skew-symmetric matrix \cite{robotics_book} (proof in Appendix \ref{appendix:system_matrices}).
\item $D$ is a symmetric and positive semi-definite matrix (follows from the definition of $D$). \\
\end{itemize}

Due to the position-dependent nature of the NAPAS setup, the LPV framework is a natural selection to represent \eqref{eqn:nlmodel} in terms of a linear model. The use of a LPV descriptor state-space representation \eqref{eqn:global_lpv_model} is more convenient as it is more closely related to the equations of motion \eqref{eqn:nlmodel}. Inversion of the mass-matrix is avoided and therefore, affine dependency in the scheduling variables is retained. The LPV representation is given as:
\begin{subequations}
\label{eqn:global_lpv_model}
\begin{align}
E(p(t))\dot{x}(t) &= A(p(t))x(t) + B(p(t))u(t) \\
y(t) &= C(p(t))x(t)
\end{align}
\end{subequations}
where $t \in \mathbb{R}^+_0$, $x(t) := \left[\begin{smallmatrix} q^\top & \dot{q}^\top \end{smallmatrix}\right]^\top \in \mathbb{R}^{n_x}$ contains the state variables, $u(t) := W \in \mathbb{R}^{n_u}$ are the control inputs, $y(t) := q \in \mathbb{R}^{n_y}$ are the outputs and $p(t) := \left[\begin{smallmatrix} p_1(t) & \cdots & p_{n_p}(t) \end{smallmatrix}\right]^\top \in \mathcal{P} \subset \mathbb{R}^{n_p}$ is the scheduling variable under the assumption that it is measurable in real time. Note that \eqref{eqn:global_lpv_model} describes a linear system with respect to the input-output (IO) partition $(u,y)$. If $p(t)$ is constant at each time instant i.e., $p(t) \equiv \bar{p}$ $\forall t \in \mathbb{R}$, then \eqref{eqn:global_lpv_model} becomes LTI, which is often referred to as the frozen dynamical description of \eqref{eqn:global_lpv_model} for a corresponding constant $p(t) \equiv \bar{p}$. If the scheduling variables $p(t)$ vary with time, the model becomes a time-varying representation. The LPV representation is said to be affine if the system matrices take an affine form i.e.,
\begin{align}
A(p(t)) = A_0 + \sum_{i=1}^{n_p} p_i(t)A_i.
\end{align}
The system matrices are:
\begin{align*}
E(p(t)) &= \begin{bmatrix} I_{n_q \times n_q} & 0_{n_q \times n_q} \\ 0_{n_q \times n_q} & M(q) \end{bmatrix} \\
A(p(t)) &= \begin{bmatrix} 0_{n_q \times n_q} & I_{n_q \times n_q} \\ 0_{n_q \times n_q} & -D-C(q,\dot{q}) \end{bmatrix} \\
B(p(t)) &= \begin{bmatrix} 0_{n_q \times n_u} \\ I_{n_q \times n_u} \end{bmatrix} \\
C(p(t)) &= \begin{bmatrix} I_{n_y \times n_q} & 0_{n_y \times n_q} \end{bmatrix}
\end{align*}
As the system description is dependent on the state variables, this representation is also called a quasi-LPV representation. Furthermore, it is a global system description as it describes the nonlinear system \eqref{eqn:nlmodel} exactly.

In the NAPAS setup, the rotational displacement of the translator is allowed to deviate slightly around the zero angle rotation (initial position in Figure \ref{fig:Translator}). Therefore, an approximation of the dynamics \eqref{eqn:nlmodel} around zero angles and angular velocities can be made by using a first-order Taylor series expansion to derive an LPV state-space representation:
\begin{subequations}
\label{eqn:local_lpv_model}
\begin{align}
\dot{x}(t) &= Ax(t) + B(p(t))u(t) \\
y(t) &= Cx(t),
\end{align}
\end{subequations}
where $t \in \mathbb{R}^+_0$, $x(t) := \left[\begin{smallmatrix} q^\top & \dot{q}^\top \end{smallmatrix}\right]^\top \in \mathbb{R}^{n_x}$ contains the state variables, $u(t) := W \in \mathbb{R}^{n_u}$ are the control inputs, $y(t):= q \in \mathbb{R}^{n_y}$ are the outputs and $p(t) := \left[\begin{smallmatrix} p_1(t) & \cdots & p_{n_p}(t) \end{smallmatrix}\right]^\top \in \mathcal{P} \subset \mathbb{R}^{n_p}$ is the scheduling variable under the assumption that it is measurable in real time. By using a Taylor series approximation, the dynamics of the model hold only locally around the linearization point. Therefore, the LPV state-space model representations is a local system description of the nonlinear system.
% ------------------------------------------------------------------------ Feedforward Methods -------------------------------------------------------------------------
\section{Feedforward control}
\label{section:ff_methods}
This section focusses on the design aspects of the feedforward methods that are taken into consideration namely, A) input nonlinearity-annihilation and decoupling, B) nonlinear feedforward, C) local LPV feedforward through inversion and D) global LPV feedforward.
%_______________________________________ Annihilation _______________________________________
\subsection{Input-nonlinearity annihilation and decoupling}
\label{section:global_annihilation}
In this section, a three-step procedure is presented. First, a pre-compensator in the input mapping is designed to account for nonlinearities. Then, another pre-compensator is designed to decouple the system. These result in a system representation that is LTI and approximately diagonal. Finally, a decoupled feedforward controller is designed for the remaining dynamics. \\
\subsubsection{Annihilation of input nonlinearities}
First we will focus on the annihilation of the input nonlinearities. Both a global a) and local b) approach are presented. The global method takes the nonlinear differential equation \eqref{eqn:nlmodel} into account. The local method is based on the linearized model \eqref{eqn:local_lpv_model}. Both methods result in annihilation of the nonlinearities present in the input. The advantage in considering the nonlinear model is that global annihilation of nonlinearities in the input is achieved, i.e., the input nonlinearities are annihilated over all trajectories of the system. When considering a locally approximated model, annihilation of nonlinearities in the input is achieved at and around the point where the linear approximation holds.

\begin{remark}
A local annihilation approach could lead to better results if model uncertainties are present. For example, a more accurate (local) model description could be obtained through system identification.
\end{remark}

\paragraph{Global input-nonlinearity annihilation}
Consider the solution of the nonlinear system \eqref{eqn:nlmodel} to be given as:
\begin{align}
\label{eqn:nl_solution}
\ddot{q} = -M^{-1}(q)\left(C(q,\dot{q})+D\right)\dot{q} + M^{-1}(q)W.
\end{align}
The goal is to find a parameter-dependent matrix $Q_1(p)$ such that a new input $\tilde{u}$ is a nonlinearity-annihilated input that satisfies the relation:
\begin{align}
\label{eqn:annihilation_equation}
W = Q_1(p)\tilde{u}.
\end{align}
Substitution of \eqref{eqn:annihilation_equation} into \eqref{eqn:nl_solution} results in a solution to the system \eqref{eqn:nlmodel} with the new input $\tilde{u}$:
\begin{align}
\label{eqn:nl_solution_annihilated}
\ddot{q} = -M^{-1}(q)\left(C(q,\dot{q})+D\right)\dot{q} + M^{-1}(q)Q_1(p)\tilde{u}.
\end{align}
By choosing
\begin{align}
Q_1(p) = M(q),
\end{align}
nonlinearity annihilation in the input mapping is achieved.

\paragraph{Local input-nonlinearity annihilation}
Consider the local model given by \eqref{eqn:local_lpv_model}. The goal is to find a matrix $Q_1(p(t))$ such that a new input $\tilde{u}(t)$ satisfies
\begin{align}
\label{eqn:eqn_ref_1}
u(t) = Q_1(p(t))\tilde{u}(t).
\end{align}
This results in an annihilation of input nonlinearities. Substitution of \eqref{eqn:eqn_ref_1} into \eqref{eqn:local_lpv_model} results in the solutions of the locally approximated system \eqref{eqn:local_lpv_model} with the new input $\tilde{u}(t)$
\begin{subequations}
\begin{align}
\label{eqn:annihilated_model}
\dot{x}(t) &= Ax(t) + B(p(t)) Q_1(p(t)) \tilde{u}(t) \\
y(t) &= Cx(t).
\end{align}
\end{subequations}
To annihilate the input parameter dependency, the following subproblem can be posed:
\begin{align}
B(p(t)) Q_1(p(t)) \tilde{u}(t) = \tilde{B}\tilde{u}(t),
\end{align}
where the static matrix $\tilde{B} \in \mathbb{R}^{n_x \times n_u}$ retains a decoupled structure. Solving this problem results in the annihilation matrix:
\begin{align}
Q_1(p(t)) = B^{\dagger}(p(t))\tilde{B},
\end{align}
where $B^{\dagger}(p(t))$ is defined as the pseudo-inverse of $B(p(t))$, which constitutes a left inverse in the case $n_x > n_u$, a right inverse if $n_x < n_u$ and an exact inverse if $n_x = n_u$. Furthermore, $B(p(t))$ should be full-rank in the scheduling region $\mathcal{P}$ i.e., the pseudo-inverse is valid in the scheduling parameter space $\mathcal{P}$. \\

\subsubsection{Decoupling}
With decoupling, we aim to find a pre-compensator $\tilde{u} = Q_2\breve{u}$ such that the system \eqref{eqn:annihilated_model}, as described by the transfer function
\begin{align}
G(s) := C(sI-A)^{-1}\underbrace{B(p(t))Q_1(p(t))}_{\tilde{B}},
\end{align}
is approximately diagonal. This amounts to choosing $Q_2$ such that
\begin{align}
G(s) = C(sI-A)^{-1}\tilde{B}Q_2
\end{align}
is approximately diagonal. Several methods are available such as dynamic decoupling, steady-state decoupling or decoupling at a specified frequency \cite{MFC}. In practice, a steady-state decoupling often suffices to achieve the desired accuracy. Therefore, the steady-state decoupling method discussed in \cite{MFC} is used here. The decoupling can be obtained by selecting the pre-compensator $Q_2 = G^\dagger(0)$, provided that $G(0)$ has full row rank. Note that, in this case, $G^\dagger(0)$ denotes the right (pseudo) inverse of $G(0)$.
\begin{remark}
Note that this decoupling method holds for LTI systems and is adapted (by annihilating the input-nonlinearities) to an LPV case. If decoupling of the nonlinear system description \eqref{eqn:nl_solution_annihilated} is required, methods such as feedback linearization can be used. See for example \cite{isidori_nonlinear_control_systems} for more details on feedback linearization.
\end{remark}

\subsubsection{Feedforward design}
The combination of annihilating input nonlinearities and decoupling enables the application of LTI feedforward methods. For the local NAPAS model, there is no coupling between states, thus we only require input-nonlinearity annihilation. Furthermore, after the annihilation of input-nonlinearities, the dynamics resemble that of a double integrator. Therefore, a mass feedforward should suffice under the rigid body assumption. By choosing $\tilde{B} = \begin{bmatrix} 0_{n_u \times n_x} & I_{n_u \times n_x} \end{bmatrix}^\top$ the feedforward controller becomes $u_\mathrm{ff}(t) = F\ddot{r}(t)$ with $\ddot{r}(t)$ being the reference acceleration and $F = I$. Including the annihilation and decoupling matrices $Q_1(p(t))$ and $Q_2$ respectively, the feedforward action becomes:
\begin{align}
u_\mathrm{ff}(t) = Q_1(p(t))Q_2F\ddot{r}(t),
\end{align}
where $Q_2 = I$.

%_______________________________________ nonlinear FF _______________________________________
\subsection{Nonlinear feedforward}
In this section, the feedforward actions are computed based upon a known reference trajectory and the nonlinear model. Given the nonlinear system represented by \eqref{eqn:nlmodel} and a desired trajectory $r$ on the generalized coordinates $q$, the inputs $W$ to steer the system over the desired trajectory should satisfy
\begin{align}
\label{eqn:nlff}
W = M(r)\ddot{r} + C(r,\dot{r})\dot{r}+D\dot{r}.
\end{align}

The system achieves perfect tracking of the reference trajectory under the following conditions \cite{robotics_book}:
\begin{itemize}
\item The trajectory $r$, $\dot{r}$ and $\ddot{r}$ exists and is known for all time.
\item The model is perfectly accurate and $q(0) = r(0)$, $\dot{q}(0) = \dot{r}(0)$ then, $q(t) = r(t)$ for all time $t \geq 0$.
\end{itemize}
This method is however not robust. In practice, the exact position of the system is not known exactly, thus a mismatch between the initial conditions of the system and reference trajectory results directly i.e., $q(0) \neq r(0)$ and $\dot{q}(0) \neq \dot{r}(0)$. The prediction of the system trajectory based upon the reference would lead to a deviation of the desired reference trajectory once disturbances are present. The feedforward does not possess an ability to correct for this mismatch. Nevertheless, an appropriate feedback controller design may be used to address this issue.
%_______________________________________ local LPV FF _______________________________________
\subsection{Local LPV feedforward}
Consider the affine local LPV model \eqref{eqn:local_lpv_model}. The objective is to find a feedforward input $u_\mathrm{ff}$ such that
\begin{subequations}
\label{eqn:ss_trajectory}
\begin{align}
\dot{x}_\mathrm{ref}(t) &= Ax(t)_\mathrm{ref} + B(p(t))u_\mathrm{ff}(t) \\
y_\mathrm{ref}(t) &= Cx_\mathrm{ref}(t)
\end{align}
\end{subequations}
where $x_\mathrm{ref}(t)$ and $y_\mathrm{ref}(t)$ denote the desired compatible state and output trajectory over time respectively. In obtaining an exact inverse of the plant, consider the concept of relative degree \cite{isidori_nonlinear_control_systems} by which the direct influence of the input $u$ can be seen through the $n$-fold derivative of the output $y$. In this work, the concept presented in \cite{yanin_exact_inverse} is used to recover dynamical relations between the desired trajectory $y_\mathrm{ref}$, state trajectory $x_\mathrm{ref}$ and feedforward input $u_\mathrm{ff}$ and their time derivatives, in the LPV setting for the case where the input matrix has affine dependency on the scheduling variables.\\

Define the $n$-th order time derivative of $y$ as
\begin{align*}
y^{[n]} := \frac{d^ny}{dt^n}
\end{align*}

Given the state and output equations in \eqref{eqn:ss_trajectory}, the $n$-th order derivative of the output $y_\mathrm{ref}$ satisfies:
\begin{align}
\label{eqn:nth-order_output_derivative}
\begin{split}
y^{[n]}_\mathrm{ref} ={} &CA^nx_\mathrm{ref} \\
&+ C\sum^{n-1}_{i=0}{A^{n-i-1} \sum^{n_p}_{j=1} B_j \sum^{i}_{k=0} \binom{i}{k}z(i,j,k) },
\end{split}
\end{align}
where $\binom{i}{k} = \frac{i!}{k!(i-k)!}$ denotes the binomial coefficient and $z(i,j,k) = p_j^{[k]}u_\mathrm{ff}^{[i-k]} + B_0 u_\mathrm{ff}^{[i]}$. Collecting the terms in \eqref{eqn:nth-order_output_derivative} dependent on the state $x_\mathrm{ref}$ and the input vector $\xi_\mathrm{ff}$ results in a more general representation:
\begin{align}
\label{eqn:nth-order_output_derivative_2}
y^{[n]}_\mathrm{ref} = \breve{E}(n)x_\mathrm{ref} + \breve{F}(n,p)\xi_\mathrm{ff},
\end{align}
with
\begin{align*}
\xi_\mathrm{ff} := \begin{bmatrix} u^\top_\mathrm{ff} & u^{\top[1]}_\mathrm{ff} & \dots &u^{\top[n-1]}_\mathrm{ff}\end{bmatrix}^\top.
\end{align*}
From \eqref{eqn:nth-order_output_derivative_2}, it can be derived that:
\begin{align}
\label{eqn:ff_equation}
\begin{split}
u_\mathrm{ff} &= \breve{C}\xi_\mathrm{ff} \\
 &= \breve{C} \breve{F}^{\dagger}(n,p)\left( y^{[n]}_\mathrm{ref} - \breve{E}(n)x_\mathrm{ref} \right),
 \end{split}
\end{align}
where $\breve{C} \in \mathbb{R}^{n_u \times n\cdot n_u}$ is a matrix that selects the feedforward input $u_\mathrm{ff}$:
\begin{align*}
\breve{C} = \begin{bmatrix}I_{(n_u \times n_u)} & 0_{(n_u \times (n-1)n_u)} \end{bmatrix}.
\end{align*}
Here,  $\breve{F}^{\dagger}(n,p)$ denotes the point-wise pseudo inverse dependent on the scheduling variable $p(t)$. If the exact or right inverse of $\breve{F}(n,p)$ exists for all scheduling variables $p(t)$, then it is possible to find the ideal $\xi_\mathrm{ff}$ that drives the system exactly according to $y^{[n]}_\mathrm{ref}$. Utilizing the system description \eqref{eqn:ss_trajectory} and \eqref{eqn:ff_equation}, the LPV system inverse is given as:
\begin{subequations}
\label{eqn:inverse_ff}
\begin{align}
\dot{x}_\mathrm{ref}(t) &= A_\mathrm{ff}(p(t))x_\mathrm{ref}(t) + B_\mathrm{ff}(p(t))y_\mathrm{ref}(t) \\
u_\mathrm{ff}(t) &= C_\mathrm{ff}(p(t))x_\mathrm{ref}(t) + D_\mathrm{ff}(p(t))y_\mathrm{ref}(t)
\end{align}
\end{subequations}
where
\begin{align*}
A_\mathrm{ff}(p(t)) &= A-B(p(t))\breve{C}\breve{F}^{\dagger}(n,p)\breve{E}(n) \\
B_\mathrm{ff}(p(t)) &= B(p(t))\breve{C}\breve{F}^{\dagger}(n,p) \\
C_\mathrm{ff}(p(t)) &= -\breve{C}\breve{F}^{\dagger}(n,p)\breve{E}(n) \\
D_\mathrm{ff}(p(t)) &= \breve{C}\breve{F}^{\dagger}(n,p).
\end{align*}
The system description \eqref{eqn:inverse_ff} is the inverse system of \eqref{eqn:ss_trajectory} that takes $y^{[n]}_\mathrm{ref}$ as input and generates the feedforward signal $u_\mathrm{ff}$ as output along the corresponding reference state trajectory $x_\mathrm{ref}$. This strategy can be used under the following conditions:
\begin{itemize}
\item the relative degree $n$ is known and it is the same for all $p \in \mathcal{P}$,
\item the initial conditions are the same: $x(t_0) = x_\mathrm{ref}(t_0)$,
\item $y^{[n]}_\mathrm{ref} \in L_2$ is known a priori with $\int_{t_0}^{\infty} \, y^{[n]}_\mathrm{ref} \, \mathrm{d}t = 0$, resulting in a bounded feedforward input,
\item $\hat{F}^{\dagger}(n,p)$ is computable in the sense of least-norm (right) or exact inverse for all parameters $p \in \mathcal{P}$ and their time derivatives $p^{[i]} \in \mathcal{P}^{[i]}$ with $i = 1,\dots,n$,
\item $p$ and the time derivatives $p^{[i]}$ are tractable in real time.
\end{itemize}
Specific to the rigid body model of the NAPAS system, the local inversion is based on the linearized model \eqref{eqn:local_lpv_model}. As the local NAPAS system only has parameter dependency in the input matrix $B(p(t))$ and the model does not contain any zeros, the relative degree $n$ is fixed at 2. The scheduling parameters $p(t)$ and their time derivatives $p^{[i]}(t)$ are all tractable as they are states of the system.

%_______________________________________ global LPV FF _______________________________________
\subsection{Global LPV feedforward}
In this section, two methods to design a global LPV feedforward controller are presented. The first method is similar to the local LPV feedforward and uses an inversion based strategy as presented in \cite{yanin_exact_inverse}. The second method is similar to the nonlinear feedforward method presented in Section \ref{section:ff_methods}. Both methods are compared through a simulation and are shown to give equivalent results. \\

\subsubsection{Global LPV feedforward through inversion}
Consider the affine global LPV model \eqref{eqn:global_lpv_model}. Similar to the local LPV inversion method, the objective is to find a feedforward input $u_\mathrm{ff}$ such that
\begin{subequations}
\label{eqn:ss_trajectory_dss}
\begin{align}
E(p(t))\dot{x}_\mathrm{ref}(t) &= A(p(t))x_\mathrm{ref}(t) + B(p(t))u_\mathrm{ff}(t) \\
y_\mathrm{ref}(t) &= C(p(t))x_\mathrm{ref}(t)
\end{align}
\end{subequations}
where $x_\mathrm{ref}(t)$ and $y_\mathrm{ref}$ denote the desired compatible state and output trajectory over time respectively. The same notion of relative degree will be used, by which the direct influence of the input $u$ can be seen through the $n$-fold derivative of the output $y$. In this section, the concept presented in \cite{yanin_exact_inverse} is used to recover dynamical relations between the desired trajectory $y_\mathrm{ref}$, state trajectory $x_\mathrm{ref}$ and feedforward input $u_\mathrm{ff}$ and their time derivatives, in the LPV descriptor state-space setting for the case where the system matrices $E(p(t))$, $A(p(t))$, $B(p(t))$ and $C(p(t))$ have affine dependency on the scheduling variables.

Given the state and output equations in \eqref{eqn:ss_trajectory_dss}, the $n$-th order derivative of the output $y_\mathrm{ref}$ satisfies:
\begin{align}
\label{eqn:dss_output_derivative}
y_\mathrm{ref}^{[n]} = \sum^{n}_{k=0} \binom{n}{k} C^{[n-k]}(p(t))x_\mathrm{ref}^{[k]}
\end{align}
where
\begin{align*}
x_\mathrm{ref}^{[n]} &= \sum^{n-1}_{k=0} \binom{n-1}{k} E_\mathrm{inv}^{[n-k-1]}(p(t))z^{[k]} \\
z^{[n]} &= \sum^{n}_{k=0} \binom{n}{k} A^{[n-k]}(p(t))x_\mathrm{ref}^{[k]}(t) + B^{[n-k]}(p(t))u_\mathrm{ff}^{[k]}(t)
\end{align*}
Let $E_\mathrm{inv}(p(t)) := E^{-1}(p(t))$ denote the inverse of $E(p(t))$ and let $\binom{n}{k} = \frac{n!}{k!(n-k)!}$ denote the binomial coefficient.
The first derivative of the inverse of a matrix $E(p(t))$ is given as \cite{matrix_cookbook}:
\begin{align*}
E^{[1]}_\mathrm{inv}(p(t)) = -E^{-1}(p(t)) E^{[1]}(p(t)) E^{-1}(p(t))
\end{align*}
and higher order derivatives of $E^{-1}(p(t))$ can be recursively computed.
The system matrices are assumed to have affine dependency on the scheduling variables, therefore the $n$-th order derivatives are given as:
\begin{align*}
E^{[n]}(p(t)) = \sum^{n_p}_{i=1}E_i p_i^{[n]}(t).
\end{align*}
Collecting the terms in \eqref{eqn:dss_output_derivative} dependent on the state $x_\mathrm{ref}$ and the input vector $\xi_\mathrm{ref}$ results in a more general representation:
\begin{align}
\label{eqn:dss_general_output_derivative}
y_\mathrm{ref}^{[n]} = \breve{E}(n,p)x_\mathrm{ref} + \breve{F}(n,p)\xi_\mathrm{ff},
\end{align}
with
\begin{align*}
\xi_\mathrm{ff} := \begin{bmatrix} u^\top_\mathrm{ff} & u^{\top[1]}_\mathrm{ff} & \dots &u^{\top[n-1]}_\mathrm{ff}\end{bmatrix}^\top.
\end{align*}
From \eqref{eqn:dss_general_output_derivative} it can be derived that:
\begin{align}
\label{eqn:dss_ff_equation}
\begin{split}
u_\mathrm{ff} &= \breve{C}\xi_\mathrm{ff} \\
 &= \breve{C} \breve{F}^{\dagger}(n,p)\left( y^{[n]}_\mathrm{ref} - \breve{E}(n,p)x_\mathrm{ref} \right),
 \end{split}
\end{align}
where $\breve{C} \in \mathbb{R}^{n_u \times n\cdot n_u}$ is a matrix that selects the feedforward input $u_\mathrm{ff}$:
\begin{align*}
\breve{C} = \begin{bmatrix}I_{(n_u \times n_u)} & 0_{(n_u \times (n-1)n_u)} \end{bmatrix}.
\end{align*}
Here,  $\breve{F}^{\dagger}(n,p)$ denotes the point-wise pseudo inverse dependent on the scheduling variable $p(t)$. If the exact or right inverse of $\breve{F}(n,p)$ exists for all scheduling variables $p(t)$, then it is possible to find the ideal $\xi_\mathrm{ff}$ that drives the system exactly according to $y^{[n]}_\mathrm{ref}$. Utilizing the system description \eqref{eqn:ss_trajectory_dss} and \eqref{eqn:dss_ff_equation}, the LPV system inverse is given as:
\begin{subequations}
\label{eqn:dss_inverse_ff}
\begin{align}
E_\mathrm{ff}(p(t))\dot{x}_\mathrm{ref}(t) &= A_\mathrm{ff}(p(t))x_\mathrm{ref}(t) + B_\mathrm{ff}(p(t))y_\mathrm{ref}(t) \\
u_\mathrm{ff}(t) &= C_\mathrm{ff}(p(t))x_\mathrm{ref}(t) + D_\mathrm{ff}(p(t))y_\mathrm{ref}(t)
\end{align}
\end{subequations}
where
\begin{align*}
E_\mathrm{ff}(p(t)) &= E(p(t)) \\
A_\mathrm{ff}(p(t)) &= A(p(t))-B(p(t))\breve{C}\breve{F}^{\dagger}(n,p)\breve{E}(n,p) \\
B_\mathrm{ff}(p(t)) &= B(p(t))\breve{C}\breve{F}^{\dagger}(n,p) \\
C_\mathrm{ff}(p(t)) &= -\breve{C}\breve{F}^{\dagger}(n,p)\breve{E}(n,p) \\
D_\mathrm{ff}(p(t)) &= \breve{C}\breve{F}^{\dagger}(n,p).
\end{align*}
The system description \eqref{eqn:dss_inverse_ff} is the inverse system of \eqref{eqn:ss_trajectory_dss} that takes $y_\mathrm{ref}^{[n]}$ as input and generates the feedforward signal $u_\mathrm{ff}$ as output along the corresponding reference state trajectory $x_\mathrm{ref}$. This strategy is valid under the following conditions:
\begin{itemize}
\item the relative degree $n$ is known and it is the same for all $p \in \mathcal{P}$,
\item the initial conditions are the same: $x(t_0) = x_\mathrm{ref}(t_0)$,
\item $y^{[n]}_\mathrm{ref} \in L_2$ is known a priori with $\int_{t_0}^{\infty} \, y^{[n]}_\mathrm{ref} \, \mathrm{d}t = 0$, resulting in a bounded feedforward input,
\item $\hat{F}^{\dagger}(n,p)$ is computable in the sense of least-norm (right) or exact inverse for all parameters $p \in \mathcal{P}$ and their time derivatives $p^{[i]} \in \mathcal{P}^{[i]}$ with $i = 1,\dots,n$,
\item $E(p(t))$ is invertible for all parameters $p \in \mathcal{P}$,
\item $p$ and the time derivatives $p^{[n]}$ are tractable in real time. \\
\end{itemize}

\subsubsection{Global LPV feedforward through input computation}
In this section, an LPV feedforward based on the global system description \eqref{eqn:nlmodel} is obtained in a similar fashion as the nonlinear feedforward. Given the nonlinear model \eqref{eqn:nlmodel} and the desired reference trajectory $r$ the input $W$ to steer the system over the desired trajectory are computed by substitution of $r$ into \eqref{eqn:nlmodel} and solving for $W$:
\begin{align}
\label{eqn:global_lpv_ff}
W = M(q)\ddot{r} + C(q,\dot{q})\dot{r} + D\dot{r}
\end{align}
The subtle difference between \eqref{eqn:nlff} and \eqref{eqn:global_lpv_ff} is the dependency of the matrices $M(q)$ and $C(q,\dot{q})$ on $q$ instead of $r$. By utilizing the actual measured state instead of reference signal, the feedforward obtains information on the current state of the system instead of relying on the information predicted by the reference trajectory. Substituting the newly computed input into the nonlinear model \eqref{eqn:nlmodel} results in the error dynamics:
\begin{align}
\label{eqn:error_dynamics}
M(q)\ddot{e} + C(q,\dot{q})\dot{e} + D\dot{e} = 0,
\end{align}
with $e := q-r$. From the error dynamics it can be concluded that the feedforward achieves perfect tracking if there is no mismatch between initial position and velocity of the system and the reference trajectory profile. The feedforward achieves accurate tracking under the following conditions:
\begin{itemize}
\item The trajectory $r$, $\dot{r}$ and $\ddot{r}$ exists and is known for all time.
\item Under the assumption that the model is accurate and $q(0) = r(0)$, $\dot{q}(0) = \dot{r}(0)$ then, $q(t) = r(t)$ for all time $t \geq 0$ (the proof follows from \eqref{eqn:error_dynamics}).
\item The pair $(q,\dot{q})$ is directly measurable in real time. \\
\end{itemize}

\subsubsection{Comparison}
\begin{figure}[h]
\begin{center}
\includegraphics[width=0.90\linewidth]{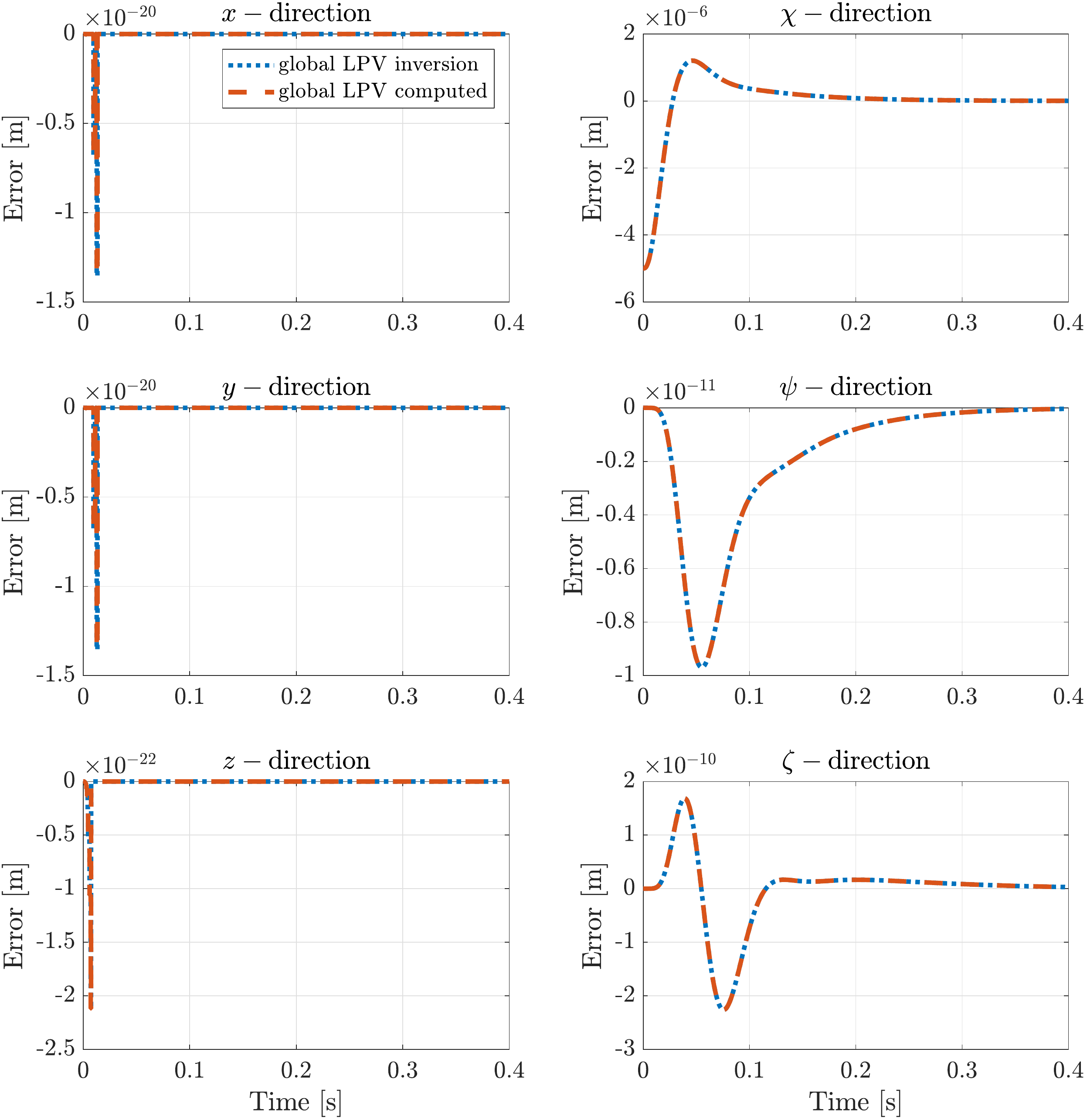}
\caption{SImulation results of the two feedforward methods, 1) global LPV feedforward through inversion and 2) global LPV feedforward through input computation, in a closed-loop simulation with an initial condition mismatch of $\chi(0) = 5$ $\mu$rad. It can be seen that the two methods have equivalent performance.}
\label{fig:ff_cl_glpv_q0}
\end{center}
\end{figure}

In this section, equivalence of the two global LPV feedforward methods, 1) global LPV feedforward through inversion and 2) global LPV feedforward through input computation, is shown through a closed-loop simulation. The closed-loop simulation environment is depicted in Figure \ref{fig:cl_block_diagram}, where the $\mathrm{FF}$-block indicates the feedforward controller, the $\mathrm{FB}$-block indicates a PID feedback controller and $r$ indicates the reference trajectory as given in Figure \ref{fig:reference_profile}. An initial condition mismatch for $\chi(0) = 5$ $\mu$rad is used for the simulation. The resulting error signal, $e:= q-r$, is shown in Figure \ref{fig:ff_cl_glpv_q0}. As expected, the two feedforward strategies perform the same.
Throughout the rest of the report, method 2) global LPV feedforward through input computation will be used as it is far more simple to implement and design than the inversion based method. Furthermore, it is less susceptible to numerical inaccuracies (e.g., caused by matrix inversions).

% ----------------------------------------------------------------------------- Stability Analysis --------------------------------------------------------------------------
\section{Stability Analysis}
\label{section:stability_analysis}
The NAPAS system is inherently unstable. Therefore, to prevent undesired behavior and possible damage to the setup, it is of importance to stabilize the system through feedback control. In this section, closed-loop stability of the NAPAS setup with the global LPV feedforward, presented in Section \ref{section:ff_methods}, and a proportional gain feedback controller is assessed. Consider the feedforward \eqref{eqn:global_lpv_ff} and feedback-controller in the form:
 \begin{align}
\label{eqn:controller}
\underbrace{M(q)\ddot{r} + C(q,\dot{q})\dot{r} + D\dot{r}}_\mathrm{feedforward} &= W + \underbrace{K_p e}_\mathrm{feedback}.
\end{align}
Here $K_p$ is a symmetric and positive definite matrix representing a proportional feedback gain, $r$ is the reference signal and $e := q-r$ is the error between the states and the reference. Substitution of \eqref{eqn:controller} in \eqref{eqn:nlmodel} yields the closed-loop dynamics:
\begin{align}
\label{eqn:error_model}
M(q)\ddot{e} + C(q,\dot{q})\dot{e} + D\dot{e} + K_p e = 0.
\end{align}
Stability of the closed-loop system is assessed by using Lyapunov's direct method. The idea follows from \cite{robotics_book}. Consider the following Lyapunov function candidate for the closed-loop dynamics \eqref{eqn:error_model}:
\begin{align}
V(\mathfrak{e},\dot{\mathfrak{e}},q) =
\frac{1}{2}
\begin{bmatrix} \mathfrak{e} \\ \dot{\mathfrak{e}} \end{bmatrix}^\top
\begin{bmatrix} K_p & \epsilon M(\mathfrak{q}) \\ \epsilon M(\mathfrak{q}) & M(\mathfrak{q}) \end{bmatrix}
\begin{bmatrix} \mathfrak{e} \\ \dot{\mathfrak{e}} \end{bmatrix} > 0,
\end{align}
which is positive for $\epsilon > 0$ sufficiently small since $M(\mathfrak{q}) \succ 0$ and $K_p \succ 0$ for values $\mathfrak{q}$ along possible trajectories of $q$. Evaluating $\dot{V}(\mathfrak{e},\dot{\mathfrak{e}},\mathfrak{q})$ along trajectories of \eqref{eqn:error_model} gives:
\begin{align}
\begin{split}
\dot{V}(\mathfrak{e},\dot{\mathfrak{e}},\mathfrak{q}) ={} &\dot{\mathfrak{e}}^\top M(\mathfrak{q})\ddot{\mathfrak{e}} + \frac{1}{2}\dot{\mathfrak{e}}^\top\dot{M}(\mathfrak{q})\dot{\mathfrak{e}} + \dot{\mathfrak{e}}^\top K_p \mathfrak{e} \\
&+ \epsilon \dot{\mathfrak{e}}^\top M(\mathfrak{q}) \dot{\mathfrak{e}} + \epsilon \mathfrak{e}^\top \left( M(\mathfrak{q}) \ddot{\mathfrak{e}} + \dot{M}(\mathfrak{q}) \dot{\mathfrak{e}} \right), \\
={} &-\dot{\mathfrak{e}}^\top\left(D-\epsilon M(\mathfrak{q})\right)\dot{\mathfrak{e}} - \epsilon \mathfrak{e}^\top\left(K_p \mathfrak{e} + D\dot{\mathfrak{e}}\right) \\
&+ \frac{1}{2}\dot{\mathfrak{e}}^\top\left(\dot{M}(\mathfrak{q})-2C(\mathfrak{q},\dot{\mathfrak{q}})\right)\dot{\mathfrak{e}}\\
&+ \frac{1}{2}\epsilon \mathfrak{e}^\top \left(2\dot{M}(\mathfrak{q}) - 2C(\mathfrak{q},\dot{\mathfrak{q}})\right)\dot{\mathfrak{e}}.
\end{split}
\end{align}
Using the skew-symmetric property of $\dot{M}(\mathfrak{q})-2C(\mathfrak{q},\dot{\mathfrak{q}})$, this results in:
\begin{align}
\dot{V}(\mathfrak{e},\dot{\mathfrak{e}},\mathfrak{q}) =
\begin{bmatrix} \mathfrak{e} \\ \dot{\mathfrak{e}} \end{bmatrix}^\top
\begin{bmatrix} -\epsilon K_p & \frac{1}{2}\epsilon (\frac{1}{2}\dot{M}(\mathfrak{q})-D) \\ \star & \epsilon M(\mathfrak{q})-D \end{bmatrix}
\begin{bmatrix} \mathfrak{e} \\ \dot{\mathfrak{e}} \end{bmatrix} < 0,
\end{align}
where $\star$ denotes terms required to make the matrix symmetric. By choosing $\epsilon > 0$ sufficiently small, we can ensure that $\dot{V}(\mathfrak{e},\dot{\mathfrak{e}},\mathfrak{q})$ is always negative along the trajectories (proof in Appendix \ref{appendix:positive_definiteness_proof}). Thus the system is asymptotically stable along trajectories. Note that asymptotic tracking requires exact cancellation of the dynamics (through feedforward) and relies on an accurate model.

\begin{remark}
If the system has no dissipation term $D$, one can apply a derivative feedback gain $K_v\dot{e}$ with $K_v$ symmetric and positive definite to achieve similar results as presented here (i.e., damping is added to the system).
\end{remark}
% ------------------------------------------------------------------------ Simulation Results ------------------------------------------------------------------------
\section{Simulation results}
\label{section:simulations}
\begin{figure}[h]
\centering
\includegraphics[width=0.70\linewidth]{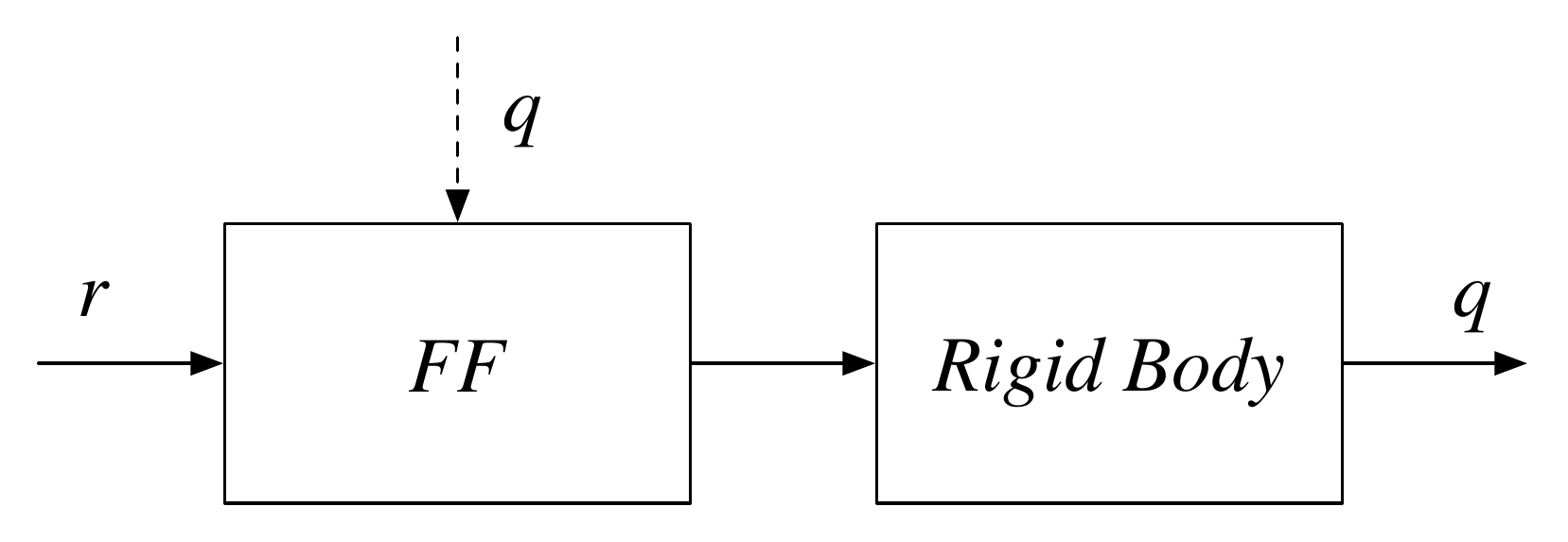}
\caption{Block diagram for the open-loop simulation. Here the $\mathrm{FF}$-block represents the feedforward controller, $r$ is the reference signal and $q$ is the system output.}
\label{fig:ol_block_diagram}
\end{figure}

\begin{figure}[h]
\centering
\includegraphics[width=0.90\linewidth]{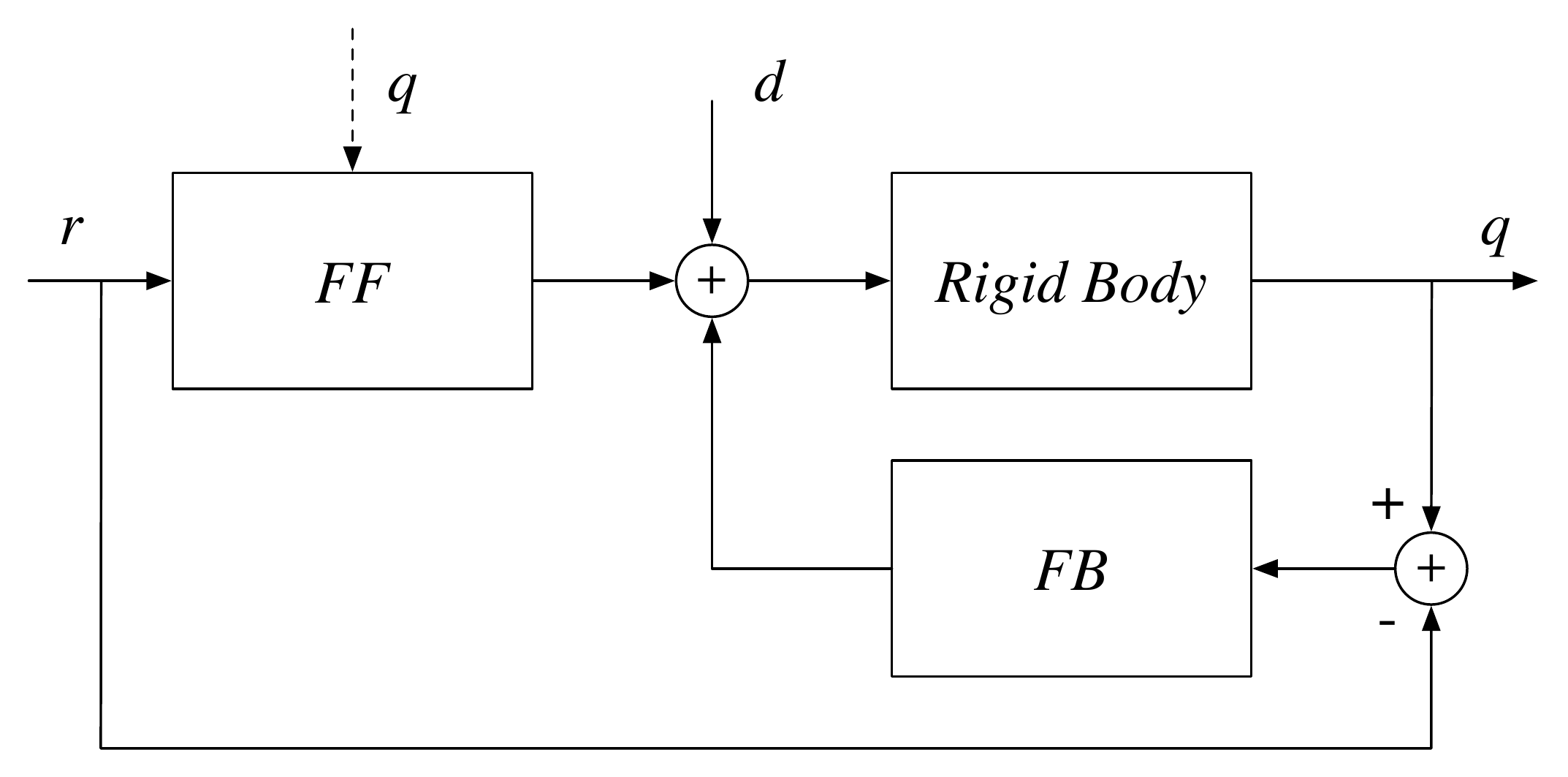}
\caption{Block diagram for the closed-loop simulation. Here the $\mathrm{FF}$-block represents the feedforward controller, the $\mathrm{FB}$-block represents the feedback controller, $r$ is the reference signal, $d$ represents an (unknown) input disturbance and $q$ is the system output.}
\label{fig:cl_block_diagram}
\end{figure}

\begin{figure}[h]
\centering
\includegraphics[width=0.90\linewidth]{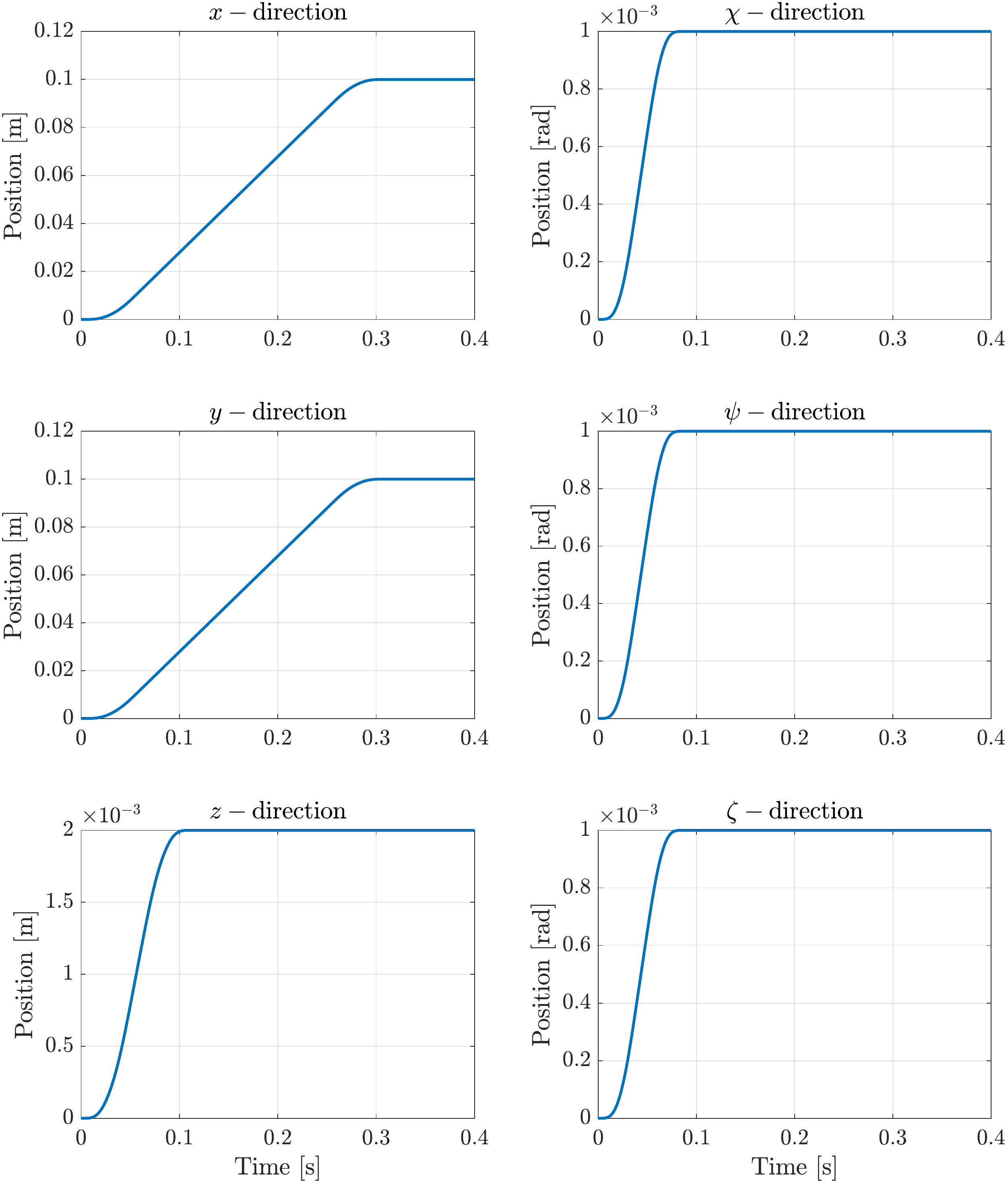}
\caption{Reference trajectory used for the simulation of the feedforward methods.}
\label{fig:reference_profile}
\end{figure}

This section provides simulation results of the four feedforward controllers, as discussed in Section \ref{section:ff_methods}, and the currently implemented mass feedforward \cite{dspace_control}. First, observations regarding the simulation results are made. This is followed by conclusions that are drawn based on the simulation data. The controllers are compared a) in open-loop simulation (Figure \ref{fig:ol_block_diagram}), to get an indication of how the feedforward controllers perform, and b) in a closed-loop simulation (Figure \ref{fig:cl_block_diagram}) to assess the performance under stable closed-loop conditions. The currently implemented decentralized SISO PID controllers, designed in \cite{dspace_control}, are used for feedback control.
The desired reference trajectory is given in Figure \ref{fig:reference_profile}.
For the open and closed-loop simulations, both matching initial conditions e.g., $r(0) = x(0)$ and an initial condition mismatch e.g., $r(0) \neq x(0)$ between the initial conditions and the reference, are simulated. The initial condition mismatch is $\chi(0) = 5$ $\mu\mathrm{rad}$ and no mismatch for the other coordinates is used.

\begin{figure}[h]
\begin{center}
\includegraphics[width=0.90\linewidth]{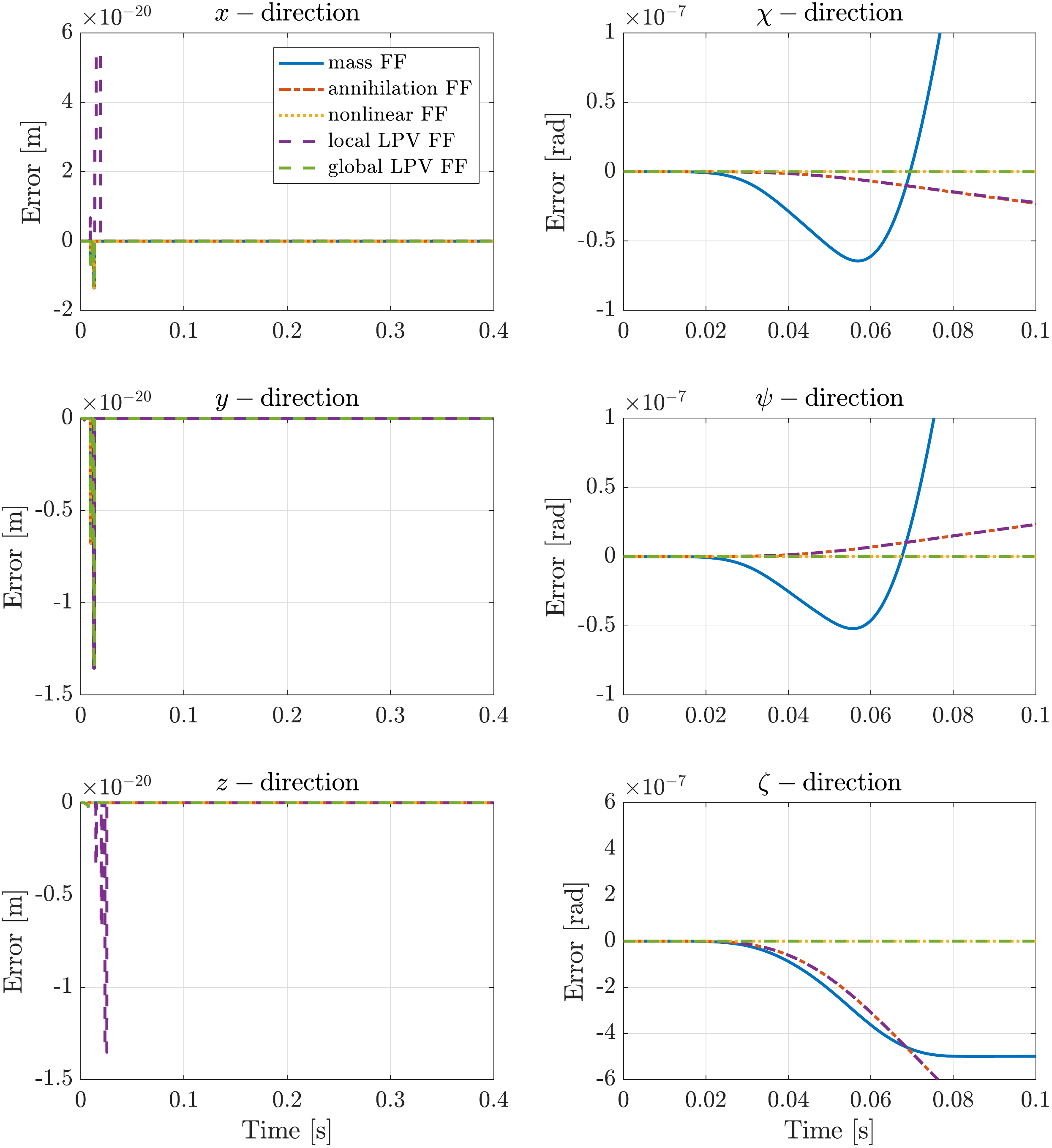}
\caption{Simulation results of the five feedforward controllers in an open-loop simulation without initial condition mismatch. Here the nonlinear and global LPV feedforward controllers perform the best.}
\label{fig:ff_ol}
\end{center}
\end{figure}

The error profiles for the open-loop simulation results, with matching initial conditions, are shown in Figure \ref{fig:ff_ol}. It is observed that all feedforward methods achieve perfect tracking for the translations. The nonlinear feedforward and global LPV feedforward methods also achieve perfect tracking for the rotations. The nonlinearity-annihilation and local LPV feedforward methods achieve better performance for the $\chi$ and $\psi$-rotations compared to the mass feedforward, whereas they perform  worse for the $\zeta$-rotation.

\begin{figure}[h]
\begin{center}
\includegraphics[width=0.90\linewidth]{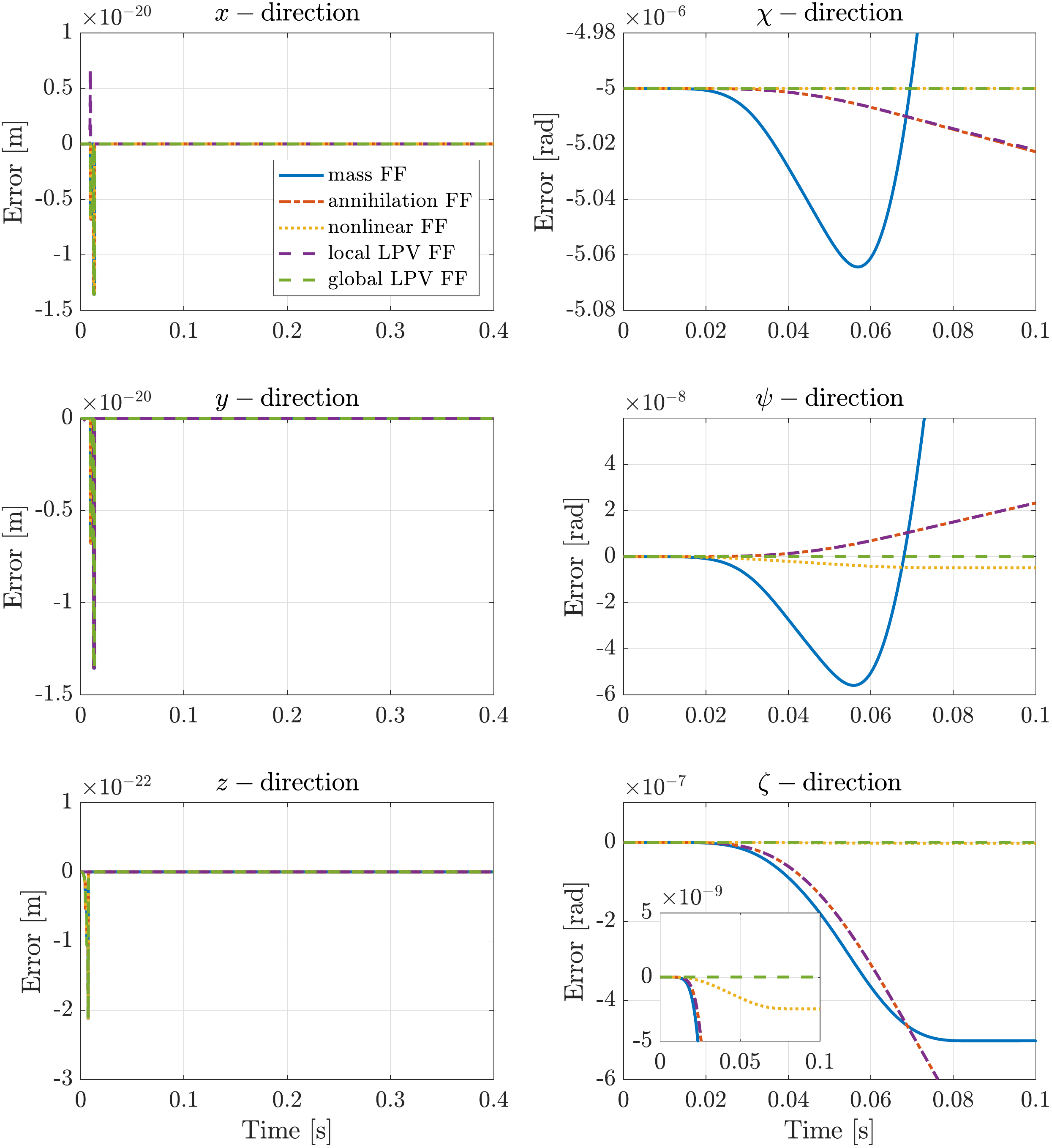}
\caption{Simulation results of the five feedforward controllers in an open-loop simulation with initial condition mismatch. Here the global LPV feedforward controller performs the best. It can be seen that the nonlinear feedforward starts to perform worse when the system deviates from the reference trajectory.}
\label{fig:ff_ol_q0}
\end{center}
\end{figure}

The error profiles for the open-loop simulation results, for an initial condition mismatch, are shown in Figure \ref{fig:ff_ol_q0}. It can be seen that an initial condition mismatch has no influence on the translations as these are still perfectly tracked. What can be observed is that the global LPV feedforward achieves perfect tracking for the $\psi$ and $\zeta$-rotations, whereas the nonlinear feedforward looses tracking accuracy.

The $\ell_2$ and $\ell_\infty$-norms of the sampled error signal, using a sampling time of $65$ $\mu\mathrm{sec}$, for the open-loop simulations are shown in Table \ref{table:data_openloop}. Here the results for the $\chi$, $\psi$ and $\zeta$-rotations are shown. The feedforward methods are indicated as follows: 1) mass feedforward, 2) input-nonlinearity annihilation and decoupling based feedforward, 3) nonlinear feedforward, 4) local LPV feedforward and 5) global LPV feedforward. Furthermore, $x(0)$ indicates matching initial conditions between the state and reference i.e., $x(0) = r(0)$ and $\cancel{x(0)}$ indicates an initial condition mismatch i.e., $x(0) \neq r(0)$. From these results, it can be seen that the local LPV feedforward performs slightly better than the input-nonlinearity annihilation method.
\begin{table}[h]
\centering
\caption{Open-loop simulation data. $\ell_2$ and $\ell_\infty$-norms of the sampled error signal with respect to the $\chi$, $\psi$ and $\zeta$ angles, using a sampling time $T_s = 65$ $\mu$sec.}
\label{table:data_openloop}
\begin{tabular}{@{}l|l|lll|lll@{}}
\toprule
OL & FF      & \multicolumn{3}{l|}{\begin{tabular}[c]{@{}l@{}}$\ell_2$-norm\\ $\times 10^{-3}$\end{tabular}} & \multicolumn{3}{l}{\begin{tabular}[c]{@{}l@{}}$\ell_\infty$-norm\\ $\times 10^{-5}$\end{tabular}} \\ \midrule
                                  &                 & $\chi$                        & $\psi$                       & $\zeta$                       & $\chi$                        & $\psi$                        & $\zeta$                       \\ \cmidrule(l){2-8}
\multirow{5}{*}{$x(0)$}    & 1)         & 0.2413                        & 0.2481                       & 0.0360                        & 0.5878                        & 0.6034                        & 0.0500                        \\
                                  & 2) & 0.0062                        & 0.0063                       & 0.2789                        & 0.0144                        & 0.0147                        & 0.6523                        \\
                                  & 3)    & 0                             & 0                            & 0                             & 0                             & 0                             & 0                             \\
                                  & 4)    & 0.0059                        & 0.0063                       & 0.2784                        & 0.0138                        & 0.0147                        & 0.6511                        \\
                                  & 5)   & 0                             & 0                            & 0                             & 0                             & 0                             & 0                             \\ \cmidrule(l){2-8}
\multirow{5}{*}{$\cancel{x(0)}$} & 1)         & 0.2525                        & 0.2478                       & 0.0362                        & 0.5064                        & 0.6029                        & 0.0503                        \\
                                  & 2) & 0.3973                        & 0.0063                       & 0.2789                        & 0.5144                        & 0.0147                        & 0.6523                        \\
                                  & 3)    & 0.3922                        & 0.0004                            & 0.0002                             & 0.5000                        & 0.0005                        & 0.0002                        \\
                                  & 4)    & 0.3971                        & 0.0063                       & 0.2784                        & 0.5138                        & 0.0147                        & 0.6511                        \\
                                  & 5)   & 0.3922                        & 0.0000                            & 0.0000                             & 0.5000                        & 0.0000                        & 0.0000                        \\ \bottomrule
\end{tabular}
\end{table}

\begin{figure}[h]
\begin{center}
\includegraphics[width=0.90\linewidth]{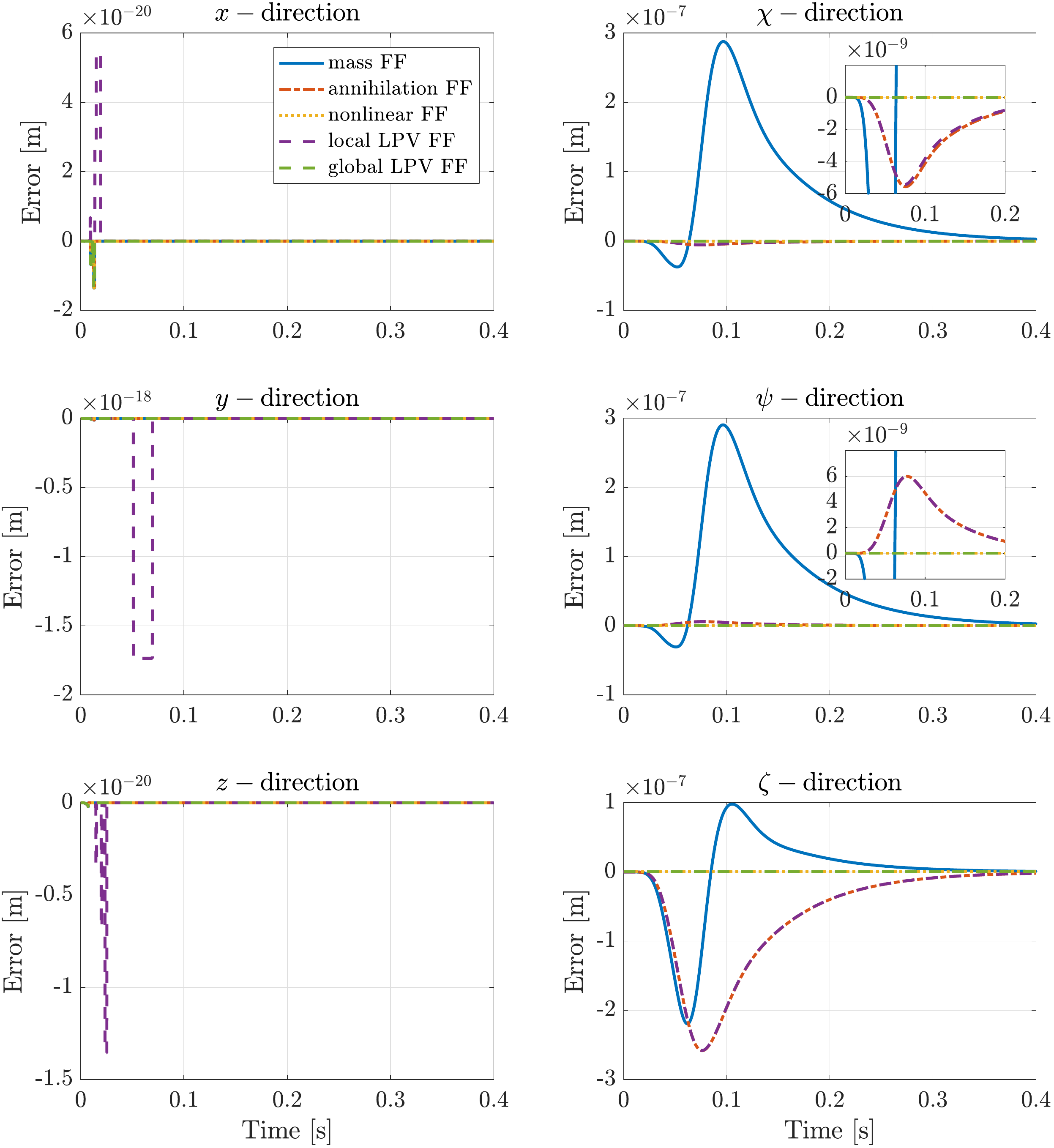}
\caption{Simulation results of the five feedforward controllers in a closed-loop simulation without initial condition mismatch. Here the nonlinear and global LPV feedforward controllers perform the best.}
\label{fig:ff_cl}
\end{center}
\end{figure}

\begin{figure}[h]
\begin{center}
\includegraphics[width=0.90\linewidth]{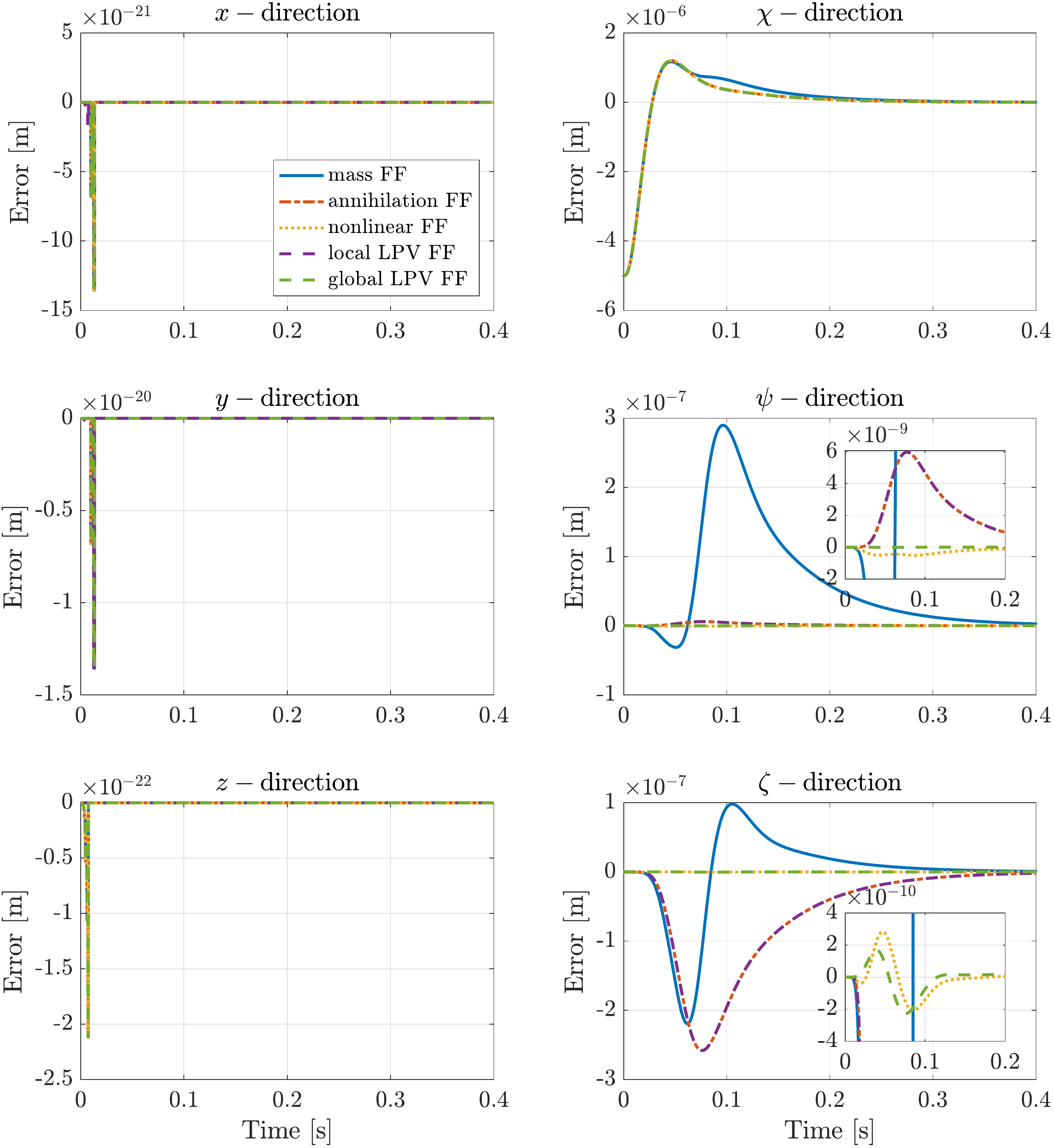}
\caption{Simulation results of the five feedforward controllers in a closed-loop simulation with initial condition mismatch. Here the global LPV feedforward controller performs the best. It can be seen that the nonlinear feedforward starts to perform worse when the system deviates from the reference trajectory.}
\label{fig:ff_cl_q0}
\end{center}
\end{figure}

The error profiles for the closed-loop simulations are shown in Figure \ref{fig:ff_cl} and Figure \ref{fig:ff_cl_q0}, for a match and mismatch in initial conditions respectively. The same observations as in the open-loop case can be made, where the slight difference between the input-nonlinearity annihilation method and local LPV feedforward become more obvious. Furthermore, the difference between the nonlinear feedforward and global LPV feedforward are more visible. This is also indicated by the $\ell_2$ and $\ell_\infty$-norms of the sampled error data, as shown in Table \ref{table:data_closedloop}.
\begin{table}[h]
\centering
\caption{Closed-loop simulation data. $\ell_2$ and $\ell_\infty$-norms of the sampled error signal with respect to the $\chi$, $\psi$ and $\zeta$ angles, using a sampling time $T_s = 65$ $\mu$sec.}
\label{table:data_closedloop}
\begin{tabular}{@{}l|l|lll|lll@{}}
\toprule
CL& FF       & \multicolumn{3}{l|}{\begin{tabular}[c]{@{}l@{}}$\ell_2$-norm\\ $\times 10^{-4}$\end{tabular}} & \multicolumn{3}{l}{\begin{tabular}[c]{@{}l@{}}$\ell_\infty$-norm\\ $\times 10^{-5}$\end{tabular}} \\ \midrule
                                  &                 & $\chi$                        & $\psi$                       & $\zeta$                       & $\chi$                        & $\psi$                        & $\zeta$                       \\ \cmidrule(l){2-8}
\multirow{5}{*}{$x(0)$}    & 1)         & 0.0803                        & 0.0812                       & 0.0486                        & 0.0288                        & 0.0290                        & 0.0220                        \\
                                  & 2) & 0.0016                        & 0.0018                       & 0.0759                        & 0.0006                        & 0.0006                        & 0.0258                        \\
                                  & 3)    & 0                             & 0                            & 0                             & 0                             & 0                             & 0                             \\
                                  & 4)    & 0.0016                        & 0.0018                       & 0.0757                        & 0.0005                        & 0.0006                        & 0.0258                        \\
                                  & 5)   & 0                             & 0                            & 0                             & 0                             & 0                             & 0                             \\ \cmidrule(l){2-8}
\multirow{5}{*}{$\cancel{x(0)}$} & 1)         & 0.7591                        & 0.0811                       & 0.0486                        & 0.5000                        & 0.0290                        & 0.0219                        \\
                                  & 2) & 0.7424                        & 0.0018                       & 0.0758                        & 0.5000                        & 0.0006                        & 0.0258                        \\
                                  & 3)    & 0.7429                        & 0.0002                       & 0.0001                        & 0.5000                        & 0.0001                        & 0.0000                        \\
                                  & 4)    & 0.7424                        & 0.0018                       & 0.0757                        & 0.5000                        & 0.0006                        & 0.0258                        \\
                                  & 5)   & 0.7429                        & 0.0000                            & 0.0001                             & 0.5000                        & 0.0000                        & 0.0000                        \\ \bottomrule
\end{tabular}
\end{table}

\begin{figure}[h]
\centering
\includegraphics[width=0.90\linewidth]{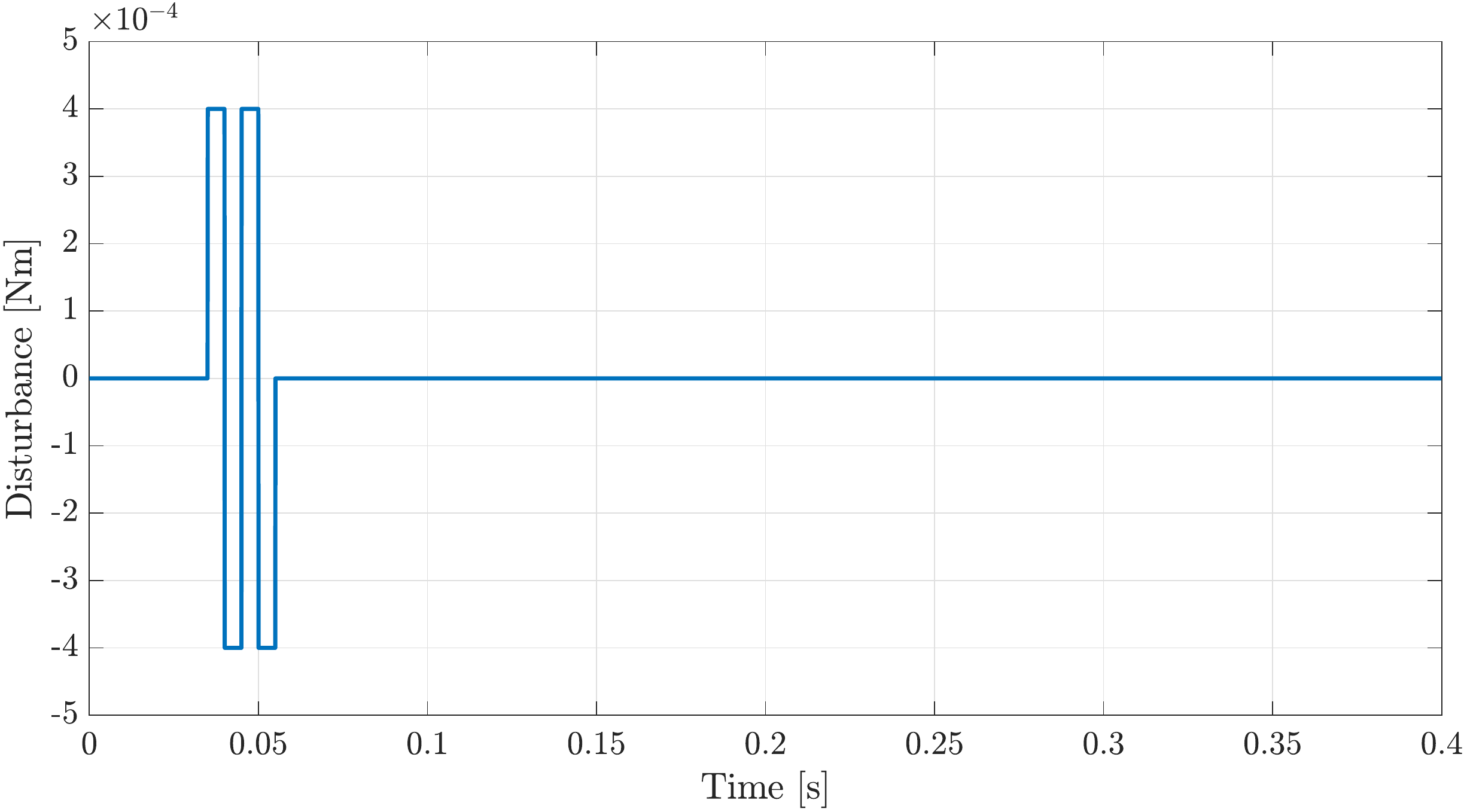}
\caption{The input disturbance profile used for the disturbance-simulation of the feedforward methods.}
\label{fig:disturbance_profile}
\end{figure}

\begin{figure}[h]
\begin{center}
\includegraphics[width=0.90\linewidth]{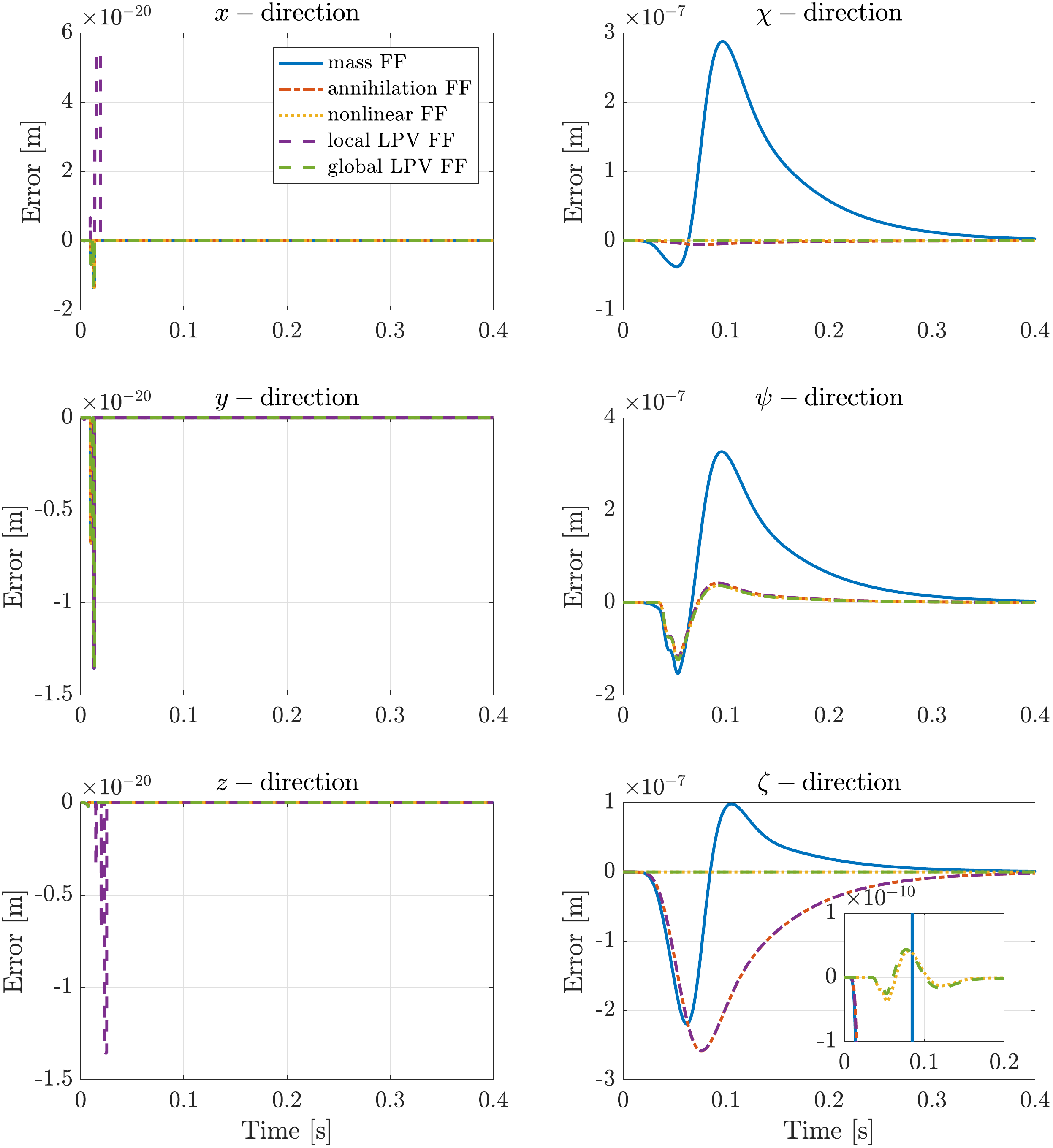}
\caption{Simulation results of the five feedforward controllers in a closed-loop simulation with an input disturbance present. Here the global LPV feedforward controller performs the best. It can be seen that the nonlinear feedforward starts to perform worse when the system deviates from the reference trajectory.}
\label{fig:ff_cl_d}
\end{center}
\end{figure}

To get an indication of how the feedforward controllers perform under disturbances, a closed-loop simulation is performed where an input disturbance is present. The disturbance profile is shown in Figure \ref{fig:disturbance_profile} and is applied on the $\tau_\psi$ input. The simulation results are shown in Figure \ref{fig:ff_cl_d}. The $\ell_2$ and $\ell_\infty$-norms of the sampled error data is shown in Table \ref{table:data_disturbance}. The results are similar as to the closed-loop simulations with initial condition mismatch. The global LPV feedforward performs the best, while the other feedforward strategies outperform the mass feedforward.
\begin{table}[h]
\centering
\caption{$\ell_2$ and $\ell_\infty$-norms of the sampled error signal with respect to the $\chi$, $\psi$ and $\zeta$ angles, using a sampling time $T_s = 65$ $\mu$sec.}
\label{table:data_disturbance}
\begin{tabular}{@{}l|l|lll|lll@{}}
\toprule
& FF       & \multicolumn{3}{l|}{\begin{tabular}[c]{@{}l@{}}$\ell_2$-norm \\ $\times 10^{-5}$\end{tabular}} & \multicolumn{3}{l}{\begin{tabular}[c]{@{}l@{}} $\ell_\infty$-norm \\ $\times 10^{-6}$\end{tabular}} \\ \midrule
                                  &                 & $\chi$                        & $\psi$                       & $\zeta$                       & $\chi$                        & $\psi$                        & $\zeta$                       \\ \cmidrule(l){2-8}
\multirow{5}{*}{III} & 1)         & 0.8031                        & 0.9234                       & 0.4862                        & 0.2877                        & 0.3267                        & 0.2196                        \\
                                  & 2) & 0.0162                        & 0.2026                       & 0.7586                        & 0.0056                        & 0.1207                        & 0.2583                        \\
                                  & 3)    & 0.0001                        & 0.2021                       & 0.0001                        & 0.0000                        & 0.1241                        & 0.0000                        \\
                                  & 4)    & 0.0155                        & 0.2026                       & 0.7573                        & 0.0054                        & 0.1207                        & 0.2580                        \\
                                  & 5)   & 0.0000                        & 0.2021                            & 0.0001                             & 0.0000                        & 0.1241                        & 0.0000                        \\ \bottomrule
\end{tabular}
\end{table}

To conclude on the results that have been presented, the translation trajectories are perfectly tracked by all feedforward methods. The plant does not have any coupling or parameter dependency regarding the translations, making even a simple mass feedforward perform excellently. However, this does not hold for the rotations of the system.

The mass feedforward does not take into account position dependent behavior of the setup. It is outperformed by all other feedforward methods except for the $\zeta$-rotation. Here it performs a bit better than the input-nonlinearity annihilation with decoupling and local LPV feedforward methods.

The input-nonlinearity annihilation method provides significant improvement for the $\chi$ and $\psi$-rotations however, the performance becomes worse for the $\zeta$-rotation in comparison to the mass feedforward. To elaborate more on this point, in the open-loop simulations for the $\zeta$-rotation, it can be seen that the trajectory goes towards negative infinity, whilst the mass-feedforward stays at a constant level. Because this method is based on a local model approximation (e.g., a linearization of the global dynamics) the model does not provide enough information, causing the decrease in performance for this rotation. Increasing the model accuracy during the design stage can improve the results of this method. However, this might introduce parameter-dependent coupling between the states, which causes difficulties in static or dynamic decoupling methods.

The nonlinear feedforward tries to cancel the dynamics of the system based on the information from the reference signal. As long as the system is exactly the same as the model and there are no disturbances, this method performs well. However, it is not very robust with respect to disturbances or model uncertainties. For example, an initial condition mismatch or disturbance has a direct effect on this method. The further away the system is from the reference trajectory, the worse the tracking performance becomes.

The local LPV inversion based-feedforward performs similarly to the nonlinearity-annihilation case. As with the nonlinearity-annihilation method, it provides significant improvements for the $\chi$ and $\psi$-rotations and a decrease in performance for the $\zeta$-rotation in comparison to the mass feedforward. Because this method is based on a local model approximation (e.g., a linearization of the global dynamics) the model does not provide enough information causing the decrease in performance for this rotation. Increasing the model accuracy during the design stage can improve the results of this method. The small difference between the nonlinearity-annihilation method and this method observed in $\chi$-rotation in Figure \ref{fig:ff_cl} is the difference in using a global nonlinearity-annihilation method (Section \ref{section:global_annihilation}) versus the local model approximation for the local LPV feedforward.
What distinguishes these methods is that the local LPV inversion can take parameter-dependent coupling between the states into account in a dynamic way. It is expected that this method will outperform the nonlinearity-annihilation method when the model accuracy increases.

The global LPV inversion can be thought of as a variation towards the local LPV inversion and the nonlinear feedforward. It describes the inverse dynamics of the nonlinear system in an LPV setting and holds globally i.e., over the entire operating range. Under perfect circumstances, it performs identically to the nonlinear feedforward and achieves perfect tracking. When disturbances or initial condition mismatches are present, it can be seen that it is more robust than the nonlinear feedforward as it takes the actual state values of the disturbed system into account instead of the prediction from the reference signals. For example, in the disturbed closed-loop simulation (Figure \ref{fig:ff_cl_q0}) the deviation from the reference in the $\zeta$-rotation, causes coupling in the $\psi$-direction which is cancelled by the global LPV inversion based feedforward. This is not cancelled by the  nonlinear feedforward. Remark, the results may vary depending on the factorization of the nonlinear model (the choice of the scheduling variables).

% --------------------------------------------------------------- Conclusions & Recommendations ---------------------------------------------------------------
\section{Conclusions and Recommendations}
\label{section:conclusions}
{The purpose of this report was to present the results of the internship on feedforward control of the magnetic levitation setup. Four feedforward methods, specifically tailored for the rigid body model of the NAPAS setup, have been developed namely: input-nonlinearity annihilation and decoupling based feedforward, nonlinear feedforward, local LPV feedforward and global LPV feedforward. These four feedforward strategies have been compared with the currently implemented mass feedforward in a simulation environment. All of the above-mentioned strategies show an increase in performance over the mass feedforward, where the global LPV feedforward has the most promising results. Furthermore, closed-loop stability of the tracking error has been assessed using the latter mentioned feedforward strategy together with a proportional feedback controller. Future investigation of the feedforward strategies on a flexible model should be investigated. Furthermore, performance of these feedforward strategies on the real world setup would be the end goal.}

\appendices
\section{System Matrices}
\label{appendix:system_matrices}
Given the NAPAS dynamics in summation form:
\begin{align*}
\sum_{j=1}^{n}M_{ij}(q)\ddot{q}_j + \sum^n_{j = 1}\sum^{n}_{k=1}\Gamma_{ijk}\dot{q}_j\dot{q}_k + \sum_{j=1}^{n}D_{ij}\dot{q}_j = W_i,
\end{align*}
for $i = 1,\cdots,n$ and $q,W \in \mathbb{R}^n$, where the mass-matrix $M(q)$ is given as
\begin{align*}
M(q) =
\begin{bmatrix}
m & 0 & 0 & 0 & 0 & 0 \\
0 & m & 0 & 0 & 0 & 0 \\
0 & 0 & m & 0 & 0 & 0 \\
0 & 0 & 0 & I_{\chi} & 0 & \alpha_2 \\
0 & 0 & 0 & 0 & I_{\psi} \cos^2(\chi) + I_{\zeta} \sin^2(\chi) & \alpha_1 \\
0 & 0 & 0 & \alpha_2 & \alpha_1 & \alpha_3
\end{bmatrix}
\end{align*}
with
\begin{align*}
\alpha_1 &= \sin(2\chi) \cos(\psi) \frac{I_{\psi} - I_{\zeta}}{2} \\
\alpha_2 &= -I_{\chi} \sin(\psi) \\
\alpha_3 &= \cos^2(\psi) \left( I_{\zeta} \cos^2(\chi) + I_{\psi} \sin^2(\chi) \right) + I_{\chi} \sin^2(\psi)
\end{align*}
The mass-matrix $M(q)$ is positive definite in the operating range, in which the angles can vary in the milliradian range.
\begin{IEEEproof}
The mass-matrix $M(q)$ is positive definite if and only if its leading leading principal minors are all positive (Sylvester's criterion). Let the determinants of the principal minors be given as:
\begin{align*}
m &> 0 \\
m^2 &> 0 \\
m^3 &> 0 \\
I_\chi m^3 &> 0 \\
I_\chi m^3\left( I_\psi \cos^2(\chi) + I_\zeta \sin^2(\chi)\right) &> 0 \\
m^3I_\chi I_\psi I_\zeta \cos^2(\psi) &> 0
\end{align*}
These principal minors are all positive if and only if
\begin{align*}
-\frac{\pi}{2} < \psi < \frac{\pi}{2}
\end{align*}
\end{IEEEproof}
Next, define the components of the coriolis matrix \\ $C(q,\dot{q}) \in \mathbb{R}^{n \times n}$ as:
\begin{align*}
C_{ij}(q,\dot{q}) &= \sum_{k=1}^{n}\Gamma_{ijk} \dot{q}_k, \quad \text{for } i,j = 1,\dots,n
\end{align*}
where $\Gamma_{ijk}$ are called the Christoffel symbols corresponding to the inertia matrix $M(q)$ and are chosen as:
\begin{align*}
\Gamma_{ijk} &= \frac{1}{2}\left( \frac{\partial M_{ij}(q)}{\partial q_k} + \frac{\partial M_{ik}(q)}{\partial q_j} - \frac{\partial M_{kj}(q)}{\partial q_i} \right).
\end{align*}
Given this formulation of the coriolis matrix, the matrix $\dot{M}(q)-2C(q,\dot{q}) \in \mathbb{R}^{n\times n}$ is a skew-symmetric matrix. The proof is taken directly from \cite{robotics_book}:
\begin{IEEEproof}
Calculate the components of the matrix $\dot{M}(q)-2C(q,\dot{q})$:
\begin{align*}
(\dot{M}(q)-2C(q,\dot{q}))_{ij} ={} &\dot{M}_{ij}(q)-2C_{ij}(q,\dot{q}) \\
={} &\sum^n_{k=1} \frac{\partial M_{ij}(q)}{\partial q_k}\dot{q}_k - \frac{\partial M_{ij}(q)}{\partial q_k}\dot{q}_k \\
&- \frac{\partial M_{ik}(q)}{\partial q_j}\dot{q}_k + \frac{\partial M_{kj}(q)}{\partial q_i}\dot{q}_k \\
={} &\frac{\partial M_{kj}(q)}{\partial q_i}\dot{q}_k - \frac{\partial M_{ik}(q)}{\partial q_j}\dot{q}_k
\end{align*}
Switching the $i$ and $j$ terms shows that $(\dot{M}(q) - 2C(q,\dot{q}))^\top = -(\dot{M}(q) - 2C(q,\dot{q}))$. Note that the skew-symmetry property depends upon the particular definition of $C(q,\dot{q})$.\\
\end{IEEEproof}
The dissipation matrix $D$ is given as:
\begin{align*}
D = \mathrm{diag}(c_1,c_2,c_3,c_4,c_5,c_6),
\end{align*}
with $c_i \geq 0$.

\section{Positive definiteness proof}
\label{appendix:positive_definiteness_proof}
\begin{IEEEproof}
Let the matrix $\mathcal{N}$ be partitioned as:
\begin{align*}
\mathcal{N}:= \begin{bmatrix}
\epsilon \mathcal{A} & \epsilon \mathcal{B} \\
\epsilon \mathcal{B}^\top & \mathcal{C} - \epsilon \mathcal{D}
\end{bmatrix} \succ 0,
\end{align*}
with $\mathcal{A}$ and $\mathcal{C}$ symmetric and positive definite. Using Schur's complement, positive definiteness of $\mathcal{N}$ is equivalent to:
\begin{subequations}
\begin{align*}
\epsilon \mathcal{A} &\succ 0, \\
\mathcal{C} - \epsilon( \mathcal{D} + \mathcal{B}^\top \mathcal{A}^{-1} \mathcal{B}) &\succ 0.
\end{align*}
\end{subequations}
Thus for $\epsilon > 0$ sufficiently small, $\mathcal{N}$ is positive definite.
\end{IEEEproof}

% Can use something like this to put references on a page
% by themselves when using endfloat and the captionsoff option.
\ifCLASSOPTIONcaptionsoff
  \newpage
\fi

% trigger a \newpage just before the given reference
% number - used to balance the columns on the last page
% adjust value as needed - may need to be readjusted if
% the document is modified later
%\IEEEtriggeratref{8}
% The "triggered" command can be changed if desired:
%\IEEEtriggercmd{\enlargethispage{-5in}}

% references section

% can use a bibliography generated by BibTeX as a .bbl file
% BibTeX documentation can be easily obtained at:
% http://mirror.ctan.org/biblio/bibtex/contrib/doc/
% The IEEEtran BibTeX style support page is at:
% http://www.michaelshell.org/tex/ieeetran/bibtex/
%\bibliographystyle{IEEEtran}
% argument is your BibTeX string definitions and bibliography database(s)
%\bibliography{IEEEabrv,../bib/paper}
%
% <OR> manually copy in the resultant .bbl file
% set second argument of \begin to the number of references
% (used to reserve space for the reference number labels box)
%\begin{thebibliography}{2}
%
%\bibitem{IEEEhowto:kopka}
%H.~Kopka and P.~W. Daly, \emph{A Guide to \LaTeX}, 3rd~ed.\hskip 1em plus
%  0.5em minus 0.4em\relax Harlow, England: Addison-Wesley, 1999.
%
%\end{thebibliography}
\bibliography{Bibliography}{}
\bibliographystyle{IEEEtran}

\end{document}